\documentclass{aa}
\usepackage[utf8]{inputenc}
\usepackage{graphicx}
\usepackage{txfonts}
\usepackage{microtype}
\usepackage{hyperref}
\usepackage{xcolor}

\makeatletter
\renewcommand*\aa@pageof{, page \thepage{} of \pageref*{LastPage}}
\makeatother

\begin{document}

\title{X-ray transients in the Chandra archive}
   \subtitle{Introducing the cumulative distribution discriminator (CuDiDi)}

   \author{I. Saathoff\inst{1}\inst{2} \and
           J. Larsson\inst{1}\inst{2}}

   \institute{KTH Royal Institute of Technology, Department of Physics, SE-10691 Stockholm, Sweden\label{inst1}
   \and
   The Oskar Klein Centre for Cosmoparticle Physics, AlbaNova University Centre, SE-10691 Stockholm, Sweden\label{inst2}\\\email{josla@kth.se}}
 
   \abstract{X-ray transients on sub-observation timescales represent a diverse and underexplored class of astrophysical phenomena, from stellar flares and magnetar bursts to extragalactic fast transients and supernova shock breakouts. We present a systematic search for such events across 20,212 \textit{Chandra} ACIS observations using a new detection pipeline that combines source identification, light-curve analysis, catalogue cross-matching, and a novel statistical classifier, the cumulative distribution discriminator (CuDiDi). From 1420 initial candidates, we identified a high-confidence golden sample of 765 transients spanning a broad range of timescales, fluxes, and spectral shapes. The candidates are distributed across the whole sky and show a wide range of durations with a median of 10~ks. A subset of fast events lasting $\lesssim$~30~s displays very soft spectra and is likely due to flaring dwarf stars, although extragalactic phenomena cannot be ruled out for all of them.  
   The comparison with previously published samples showed that CuDiDi identifies most known transients while imposing somewhat stricter variability criteria, and it also extends the total sample of \textit{Chandra} transients to include shorter events. We deliver a comprehensive catalogue of sub-observation \textit{Chandra} X-ray transients and establish a general method for exploiting archival datasets to uncover rare short-lived high-energy phenomena.
}
   

   \maketitle
   
\section{Introduction}\label{sec:introduction}

The X-ray sky is home to a rich variety of transient phenomena, ranging from magnetic reconnection in stellar atmospheres to tidal disruption events (TDEs) and various cataclysmic explosions. The importance of X-ray observations of these phenomena is highlighted by the results from the recently launched \textit{Einstein Probe} (EP; \citealt{Yuan2025}), which has opened up a new discovery space for X-ray transients. Observations with the EP have revealed a large diversity in the X-ray emission that accompanies the explosive deaths of massive stars as supernovae (SNe) and gamma-ray bursts (GRBs), providing new insights into the nature of the central engines and the properties of relativistic jets.  
This diversity is illustrated by detailed studies of individual EP transients, including cases associated with GRBs (EP240315a, \citealt{Liu2025}; EP240801a, \citealt{Jiang2025}; EP241107a, \citealt{Eappachen2026}), collapsars (EP240414a, \citealt{Sun2025,vanDalen2025}; EP250108a, \citealt{Rastinejad2025}; EP250827b, \citealt{Gokul2025}), and a likely binary neutron star merger (EP250207b, \citealt{Becerra2026,Jonker2026})
 
At the same time, archival data from X-ray telescopes 
operating since the early 2000s 
provide a rich resource for identifying transients serendipitously caught within the field of view (FOV) of other targets. Of particular interest are data from \textit{XMM-Newton} \citep{Jansen2001} and \textit{Chandra} \citep{2000SPIE.4012....2W}, both launched in 1999, as well as \textit{Swift} \citep{Burrows2005}, launched in 2004, which all cover the soft X-ray band ($\sim$ 0.5--10~keV). A number of searches for transients in the archives of these telescopes have been presented, including \cite{2020ApJ...896...39A}, \cite{Pastor-Marazuela2020}, \cite{DeLuca2021}, \cite{Ruiz2024}, and \cite{Khan2025} for \textit{XMM-Newton}, \cite{2019MNRAS.487.4721Y}, \cite{Lin2022}, \cite{2022A&A...663A.168Q}, \cite{2023A&A...675A..44Q}, \cite{2023MNRAS.523.2513Z} and \cite{Dillmann2025} for \textit{Chandra}, and \cite{EylesFerris2022} for \textit{Swift}.
Since the transients detected in these searchers are discovered years to decades after they occurred, it has not been possible to perform multi-wavelength follow-up observations, although we note that the field of the \textit{Chandra} fast X-ray transient (FXT)~141001 was serendipitously observed at optical wavelengths with the Very Large Telescope VIsible MultiObject Spectrograph only 80~min after it occurred. The observation did not reveal any counterpart \citep{2017MNRAS.467.4841B}. 
There are now efforts to identify X-ray transients from \textit{Swift} and \textit{XMM-Newton} in real time to enable follow-up observations \citep{Evans2023,Quintin2024}.

The transients identified in the archival searches have uncovered diverse phenomena, ranging from flares from Galactic objects (e.g. \citealt{Khan2025}) to a population of energetic extragalactic transients referred to as  FXTs, some of which may be powered by millisecond magnetars created in binary neutron star mergers \citep{2017MNRAS.467.4841B,Xue2019,Lin2022,2022A&A...663A.168Q,2023A&A...675A..44Q,QuirolaVasquez2024,QuirolaVasquez2025}. Adding to the diversity of X-ray transients, \citet{2020ApJ...896...39A} identified events with properties consistent with SN shock breakouts (SBOs).  This included the case of FXT~110621 (also identified by \citealt{Novara2020}), whose properties are similar to the X-ray SBO from SN2008D, for which the connection between the X-ray signal caught by \textit{Swift} and the optical SN could be firmly established \citep{Soderberg2008}. 

Although multi-wavelength follow-up observations of transients discovered in archival data are not possible, the identification and studies of their host galaxies can provide insights into their origin by constraining their luminosities and environments.
A number of such studies have been carried out \citep{Novara2020,Lin2022,Eeappachen2022,Eappachen2023,Inkenhaag2024,Eappachen2024,QuirolaVasquez2025}, but in many cases, there are either multiple host candidates or no detected host galaxy \citep{Lin2022,Eeappachen2022,Eappachen2023,Inkenhaag2024}. 
A handful of the FXTs identified in \textit{Chandra} observations are associated with very nearby galaxies ($\sim$4--30~Mpc), while the majority are more distant, with redshifts $z \gtrsim 0.1$ \citep{2022A&A...663A.168Q,2023A&A...675A..44Q}. The transients uncovered in \textit{XMM} observations by \cite{2020ApJ...896...39A} whose host galaxies are identified also fall into the latter category \citep{Novara2020,Eappachen2024}.
The FXTs with clear host galaxy associations reveal a wide range of peak luminosities, ranging from $\sim 10^{38}-10^{40}~\rm{erg\ s^{-1}\ cm^{-2}}$ for the nearby subset \citep{2022A&A...663A.168Q} to between $\sim 2 \times 10^{43}~\rm{erg\ s^{-1}\ cm^{-2}}$ (FXT~110621,\citealt{Novara2020,Eappachen2024}) and $\sim 3 \times 10^{47}~\rm{erg\ s^{-1}\ cm^{-2}}$ (FXT~141001, \citealt{QuirolaVasquez2025}) for the distant cases. This indicates multiple progenitor channels.

The transients are also diverse in terms of timescales, spectra, and fluxes, which presents a challenge for identifying them in archival searches.  
The situation is further complicated by the inhomogeneous nature of the archival data, including large variations in the duration of individual observations. The different transient searches above have therefore uncovered transients with partly different properties, depending on the datasets and detection algorithms used.

We present a new transient detection algorithm that we applied to the full public archive of the \textit{Chandra} advanced CCD imaging spectrometer (ACIS; \citealt{Garmire2003}). This transient search was designed to further exploit the potential of the archival data by also capturing faint 
and short transients, as well as transients occurring in observations with short durations. 
{The durations of many EP transients are $\sim$100-1000~s (e.g. \citealt{Sun2025,Liu2025,Zhou2026}), down to $\sim 12$~s for EP240408a \citep{Zhang2025}, while the durations of the transients so far uncovered in the \textit{Chandra} archive are  $\sim$3--50~ks \citep{2022A&A...663A.168Q,2023A&A...675A..44Q}. This suggests that a population of short \textit{Chandra} transients might not have been identified yet.

To identify robust candidates, we developed a multi-stage pipeline involving automated source detection, light-curve analysis, and catalogue cross-matching, followed by a novel statistical classification based on the cumulative distribution of the total light-curve counts. A final round of visual screening was used to flag artefacts and define a golden sample of high-confidence transients. The full pipeline and data selection process are described in Sect.~\ref{sec:identifying-transient-candidates}. In Sect.~\ref{sec:properties} we present the X-ray properties of the golden sample. This is followed by 
a discussion in Sect.~\ref{sec:discussion}, covering the properties of a small population of very short transients and an assessment of the algorithm performance. The paper concludes with a summary in Sect.~\ref{sec:summary}.

\section{Identifying transient candidates}\label{sec:identifying-transient-candidates}

An X-ray source detected as a transient in a given observation may be either a true transient in the sense that it happens only once, or be a flare from a source that often also emits faint persistent emission together with occasional bright flares. The former category includes GRBs originating from collapsars and binary neutron star mergers \citep{Kumar2015,Sun2025,Liu2025, Jonker2026},
 SBOs \citep{Waxman2017,Levinson2020,2020ApJ...896...39A}, and TDEs \citep{2021SSRv..217...40R}, where the type of TDE most relevant for producing short transients is the disruption of a white dwarf by an intermediate mass black hole \citep{Jonker2013,Maguire2020,LiD2025,Levan2025}. 

The category of sources with flaring activity in the X-ray band include active galactic nuclei (AGNs; 
\citealt{Grupe2001,2022MNRAS.510.4063S,Liu2025A}), X-ray binaries (XRBs, \citealt{Galloway2020,Sguera2020,Romano2023,Haberl2025}),
magnetars \citep{2006Woods_Thompson,2011ASSP...21..247R,Kaspi2017}, cataclysmic variables/novae \citep{Starrfield2016,2022Natur.605..248K}, and stars \citep{Pye2015,DeLuca2020,2024LRSP...21....1K,Zhao2026}.
We focused on identifying true transients and new flaring objects, meaning that we excluded transients associated with known X-ray sources. We also excluded transients associated with stars.  

\subsection{Sample selection}

We used data from the \textit{Chandra}/ACIS archive, which is well suited for identifying faint transients due to the combination of high angular resolution and a low background. ACIS provides imaging spectroscopy in the 0.3--8~keV energy range 
and comprises ten CCDs, where the four ACIS-I chips are arranged in a square covering $17\arcmin\times17\arcmin$ and the six ACIS-S chips are arranged in a row covering $50\arcmin\times8.3\arcmin$.\footnote{\url{https://cxc.harvard.edu/cal/Acis/}} Up to six (I and S) chips can be active in a given observation, which sets the maximal FOV. 

All publicly available ACIS observations obtained between 1999-08-14 and 2024-12-31 were retrieved from the high energy astrophysics science archive research center (HEASARC) using the \texttt{chanmaster} mission interface.\footnote{The data were queried via the \texttt{astroquery.heasarc} module in Python, using the \texttt{chanmaster} mission catalogue.} There are a total of 21,914 observations,
out of which 20,212 had data available for download. 
The total exposure time of these observations is 467~Ms, where 63~\% (37\%) has ACIS-S (ACIS-I) as the aimpoint. The sample includes all ACIS operating modes, including various subarrays and observations with gratings. 

The necessary event and auxiliary files (\texttt{evt2, asol, bpix, msk}) were downloaded for each observation to enable the extraction of images, light curves, and spectra. The data were not reprocessed because the benefits are small for most observations, while the computational cost is high.
We verified this by reprocessing $\sim 100$ randomly selected observations and comparing the results, which showed minimal effect on transient selection. We did note small differences in the spectra in some cases, but this will not significantly impact our objectives  to identify transients and characterize their basic properties.  
The flow chart in \autoref{fig:flowchart} shows an overview of the analysis procedure applied to each observation, and the full details are provided below.

\subsection{Source detection and extraction of science products}

\begin{figure*}[htbp]
	 \includegraphics[width=\textwidth]{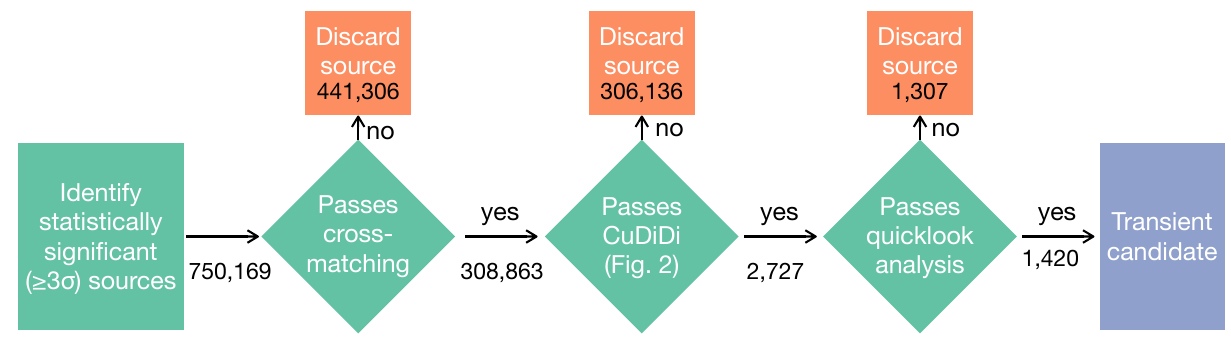}
    \centering
	 \caption{Flow chart of the transient candidate search method. The numbers shown at each step indicate how many sources were retained and excluded at that stage of the analysis.
  }
	 \label{fig:flowchart}
\end{figure*}

Images were generated using the \textit{Chandra} interactive analysis of observations (CIAO; \citealt{2006SPIE.6270E..1VF}, version 4.15.2 with CALDB 4.10.7) tool \texttt{fluximage}, restricted to the 0.3--2~keV energy range to optimise the detection of soft X-ray sources. 
This energy range was chosen to optimise the search for SBOs in future analysis of the selected transients, as these are expected to have very soft spectra \citep{Waxman2017}.   
The encircled counts fraction was set to 0.393, corresponding to the 1$\sigma$ integrated volume of a 2D Gaussian. Source detection was performed using \texttt{wavdetect}, with wavelet scales of \texttt{[1 2 4 6 8 12 16 24 32]} pixels to cover a broad range of spatial scales, and a detection threshold of $10^{-6}$. 
A total of 1,019,308 sources were detected in all the \textit{Chandra} observations, out of which 750,169 had a significance $>3\sigma$ (as reported by \texttt{wavdetect}) and were retained for further analysis. The sources detected at lower significance were used in a later stage of the analysis to flag possible recurring sources among the final transient candidates (Sect.~\ref{sec:flags}). 

The light curves were extracted using \texttt{dmextract} in the 0.5--7~keV energy range. The source regions were defined as the $3\sigma$ elliptical output regions of \texttt{wavdetect}, excluding 
any areas that overlapped with the $3\sigma$ regions of nearby sources. 
The background counts were extracted from annular regions with inner and outer radii of $2\,r_\text{max}$ and $3\,r_\text{max}$, respectively, where $r_\text{max}$ is the semi-major axis of the source ellipse. As for the source regions, any areas of background regions that overlapped with nearby sources were excluded. Point sources that overlapped with extended X-ray sources were flagged at a later stage of the analysis (Sect.~\ref{sec:flags}).

Although the nominal \textit{Chandra} ACIS frame time is 3.2~s, a 1~s binning was adopted for the cumulative distribution analysis to provide consistent temporal sampling across all observations, 
including observations that have a better time resolution due to the use of subarrays. 
For reference, 21~\% of the observations in our sample have a time resolution $< 3$~s and 13~\% have a time resolution $< 1$~s. The choice of 1~s bins has a negligible impact on the CuDiDi selection (see Sect.~\ref{sec:selection}). 
Good-time intervals (GTIs), which exclude periods of high background due to flaring or instrumental artefacts, were applied to all light curves, and the same GTIs were used for the spectral extraction. These GTIs were generated following standard CIAO tools and filtering criteria.\footnote{See \url{https://cxc.cfa.harvard.edu/ciao/ahelp/dmextract.html} for details.}

For spectral analysis, source spectra were extracted directly from the event files using \texttt{specextract}, using the same source and background regions as for the light curves. CIAO generates background-subtracted spectra along with the corresponding response matrix files and ancillary response files for each source. The spectra were grouped to 1 count per spectral bin, needed for fitting with C-stat in XSPEC, and the point spread function (PSF) correction was applied. In cases where the standard background region resulted in extraction errors, larger annuli (up to $12\,r_\text{max}$) were used to ensure a valid background spectrum. These spectral products are used for quick-look inspection and subsequent spectral fitting.
A full list of tools and parameters used throughout the pipeline is provided in \autoref{tab:pipeline parameters}.

\begin{table*}
\caption{Overview of software tools, input parameters, and selection criteria.}
\label{tab:pipeline parameters}
\centering
\begin{tabular}{lll}
	\hline\hline
	Level & Tool/Criterion & Parameters \\ 
	\hline\\
	Observation & \texttt{download\_chandra\_obsid} & \texttt{evt2, asol, bpix, msk} \\
         & \texttt{fluximage} & energy range 0.3 - 2~keV \\
         & & encircling counts fraction = 0.393 \\
         & \texttt{wavdetect} & \texttt{[1 2 4 6 8 12 16 24 32]} (default)\\
         & & sigthresh = $10^{-6}$ (default) \\
         & & $\geq 3\sigma$ significance threshold \\
         \\\hline\\
         Sources & \texttt{dmextract} & energy range 0.5-7~keV \\
         & & 3$\sigma$ region \texttt{wavdetect}, bkg annulus (2-3)r$_\text{max}$ \\
         & \texttt{specextract} & 3$\sigma$ region \texttt{wavdetect}, bkg annulus (2-3)r$_\text{max}$ \\
         \\\hline\\
         Criteria & NED cross-matching & 3\arcsec\ radius; excluding star-like types \\ 
         & SIMBAD cross-matching & 3\arcsec\ radius; excluding star-like object types \\ 
         & Gaia cross-matching & 3\arcsec\ radius \\ 
         & STONKS cross-matching & 3\arcsec\ radius \\ 
         & eRASS1 cross-matching & 3\arcsec\ radius \\
         & Cumulative distribution & $\mu_\text{upper} > \mu_\text{lower} + 0.5 + 0.2$ \\
	\\\hline
\end{tabular}	
\end{table*}

\subsection{Selection criteria}
\label{sec:selection}

Each detected source was then required to pass a series of selection criteria to be classified as a transient candidate. The criteria are as follows: 
(a) cross-matching with external catalogues to remove known sources;
(b) application of the CuDiDi criterion; and
(c) visual inspection via quick look analysis.
Each step is described in detail below.

\subsubsection{(a) Cross-matching}
Many sources identified by the \texttt{wavdetect} algorithm are expected to 
be known persistent or recurring sources, or associated with stars
\citep[see e.g.][]{2019MNRAS.487.4721Y}. To identify and exclude such sources, cross-matching was performed using a 3\arcsec\  matching radius against several astronomical catalogues. This radius was chosen to account for the combined astrometric uncertainties of the \textit{Chandra} source positions and the comparison catalogues. While \textit{Chandra} offers sub-arcsecond astrometric precision for on-axis sources (typically $\lesssim$ 0.6\arcsec), the PSF broadens significantly at larger off-axis angles, leading to increased positional uncertainties that can exceed 1\arcsec\ \citep{2024ApJS..274...22E}.\footnote{See also \url{https://cxc.harvard.edu/cal/ASPECT/celmon/}.} Additionally, small systematic offsets between different datasets may be present. The 3\arcsec\ radius thus provides a conservative but effective compromise.

For reference, out of the 750,169 sources detected at $>3 \sigma$ significance with \texttt{wavdetect}, $\sim 38,000$ ($\sim 5$\% ) have $1\sigma$ positional uncertainties $>1$\arcsec, while only 27 sources have $1\sigma$ positional uncertainties $> 3$\arcsec. Assuming that the positional probability distributions are described by circular 2D Gaussians, we estimate the number of sources that will have their true positions outside the 3\arcsec\ matching radius as
$N_{\rm outside} = \sum_{\rm i=1}^{N} e^{-3^2/(2\sigma_{\rm i}^2)} \approx 3,300$, where $\sigma_i$ is the positional uncertainty of each source and the sum is over all N detected sources. This corresponds to 0.43~\% of the sample and shows that the probability of missing real counterparts is very low, although we note that the positional errors in comparison catalogues will add to the uncertainties. The probability of excluding genuine new sources due to mismatches is higher, as discussed further below.   

First, matches were sought in the SIMBAD database \citep{2000A&AS..143....9W}, where we filter out sources categorised under the “Taxonomy of Stars” and “Sets of Stars” classifications.
Sources without a counterpart in SIMBAD were then cross-matched against the NASA/IPAC extragalactic database (NED), excluding objects classified as stellar, planetary nebulae, or SNe.\footnote{See \url{https://ned.ipac.caltech.edu/help/ui/nearposn-list_objecttypes?popup=1} for a full list.} We verified that there were no X-ray sources detected shortly before the optical detections of SNe, implying that this cut did not exclude any SBO candidates. Finally, unmatched sources were cross-matched with Gaia \citep{2016A&A...595A...1G} data release 3 (Gaia DR3; \citealt{2023A&A...674A...1G}) to identify stellar counterparts, ensuring that stars not already flagged in SIMBAD or NED were excluded. The Gaia cross-match helps to exclude nearby stars that may show low-level variability or flares. The relatively large matching radius of 3\arcsec\ also ensures that stars with significant proper motion are excluded, although we note that rare cases of stars with very high proper motion may have been missed.

Additional cross-matches were performed with the STONKS catalogue \citep{Quintin2024}, which compiles known X-ray sources across a broad range of timescales (including data from \textit{XMM-Newton}, \textit{Swift}, \textit{Chandra}, \textit{eROSITA}, and \textit{ROSAT}; see \citealt{Quintin2024} for details), and the \textit{eROSITA} all-sky survey first data release (eRASS1)  \citep{2024A&A...682A..34M}, which lists X-ray sources detected in the 0.2--2.3~keV band with detection likelihoods greater than 6.
In cases where the only match in STONKS was with a source in the \textit{Chandra} source catalog (CSC), we searched the latest version of the catalogue (CSC 2.1, \citealt{Evans2024}) and our own catalogue to exclude sources that were detected at $>3\sigma$ in more than one observation. As the version of the CSC included in STONKS  only contains observations up to the year 2014 (CSC 2.0), we used our own catalogue of detections to identify and exclude recurring sources in later observations. 
These cross-matches serve to remove known and recurring X-ray emitters from our sample, such as AGN and XRBs. 
The sample could potentially be further cleaned from flaring stars by using the various observed correlations for stellar flares in X-rays (e.g. \citealt{Pye2015}), although this is likely to provide a clear diagnostic only for the brightest cases.

In addition to the catalogue cross-matching, we manually compared our candidates with the transient samples reported by \citet{2022A&A...663A.168Q, 2023A&A...675A..44Q} (QV transients). Out of the 22 published QV transients, we recovered 13 in our analysis and removed them from our catalogue. The remaining 9 sources were not retained, either because they were associated with known objects in SIMBAD or NED, or did not pass our CuDiDi threshold. A detailed overview is provided in \autoref{tab:QVcomparison} (\autoref{app:QVcomparison}). This is also discussed further in Sect.~\ref{sec:algoperformance}.

In total, the cross-matching described above removed 441,306 out of 750,169 sources from the sample. 
The majority of the excluded sources (70\%), were removed in the cross-matching with SIMBAD, NED, and Gaia, while the rest were removed in the X-ray cross-matching. 

The conservative matching radius that was used to minimise contamination from known and recurring sources comes at the expense of a higher risk of excluding genuine transients due to matches with nearby unrelated sources. This is exemplified by the case of CDSF-S~XT2 (FXT 150322, \citealt{Xue2019}), which was initially misidentified with a nearby AGN. The probability of a mismatch will be different for each source, depending on the source density in the field, as well as its positional uncertainty and uncertainties in the matching catalogues. To obtain a rough estimate of the impact of mismatches, we reran the matching step for 500 randomly selected observations (containing $\sim 19,000$ sources) using a more conservative matching radius of 1.5\arcsec. The $3\sigma$ positional uncertainties on the \textit{Chandra} sources are smaller than this for 80\% of the sample, so matches with more distant sources have a higher probability to be due to unrelated nearby sources. 

We found that the more conservative radius reduces the matches in the SIMBAD/NED/Gaia step by 27\%, while the X-ray matches are reduced by 14\%. The smaller difference in the latter case is partly due to matches with recurring \textit{Chandra} sources, and when these matches were excluded, the reduction in the X-ray cross-matching was 22\%.
While this investigation indicates that the probability of mismatches is relatively high, we note that this conservative approach also implies that the analysis of the final selected transients is unlikely to be affected by contamination of nearby sources.

\subsubsection{(b) Cumulative distribution discriminator}
The CuDiDi is a novel statistical tool we developed to identify transient sources based on the shape of their light-curve distributions (see \autoref{fig:cudidi}). While the method is in principle sensitive to repeating and persistent behaviours as well, its design and application here are focused specifically on isolating transients.

For each source, the cumulative distribution function (CDF) of the photon arrival times is calculated from the 0.5--7~keV energy range light curves, normalised and divided at the median (y=0.5) value of the CDF into two halves: the lower half (0--0.5) and the upper half (0.5--1.0). The mean values of each half are computed and denoted as $\mu_{\text{lower}}$ and $\mu_{\text{upper}}$, respectively.
Background counts, which are usually evenly spread over time, can slightly flatten the shape of the CDF, reducing the contrast between the two halves and affecting the calculated mean arrival times ($\mu_{\text{lower}}$ and $\mu_{\text{upper}}$). Subtracting the background is not practical in our case because it can produce negative counts in bins with low signal, complicating the analysis. However, since \textit{Chandra} generally has low background levels, this effect is expected to be minimal in most cases
(see \autoref{app:bkg} for further details.)

The CuDiDi diagram (\autoref{fig:cudidi}) plots $\mu_{\text{lower}}$ versus $\mu_{\text{upper}}$ for each source, with classification boundaries defined by a diagonal band centred on $\mu_{\text{upper}} = \mu_{\text{lower}} + 0.5$, with a width of $\pm 0.2$. Sources above this band are classified as transients, those within it as persistent, and those below as repeating.
\begin{figure*}[htbp]
	 \includegraphics[width=.8\textwidth]{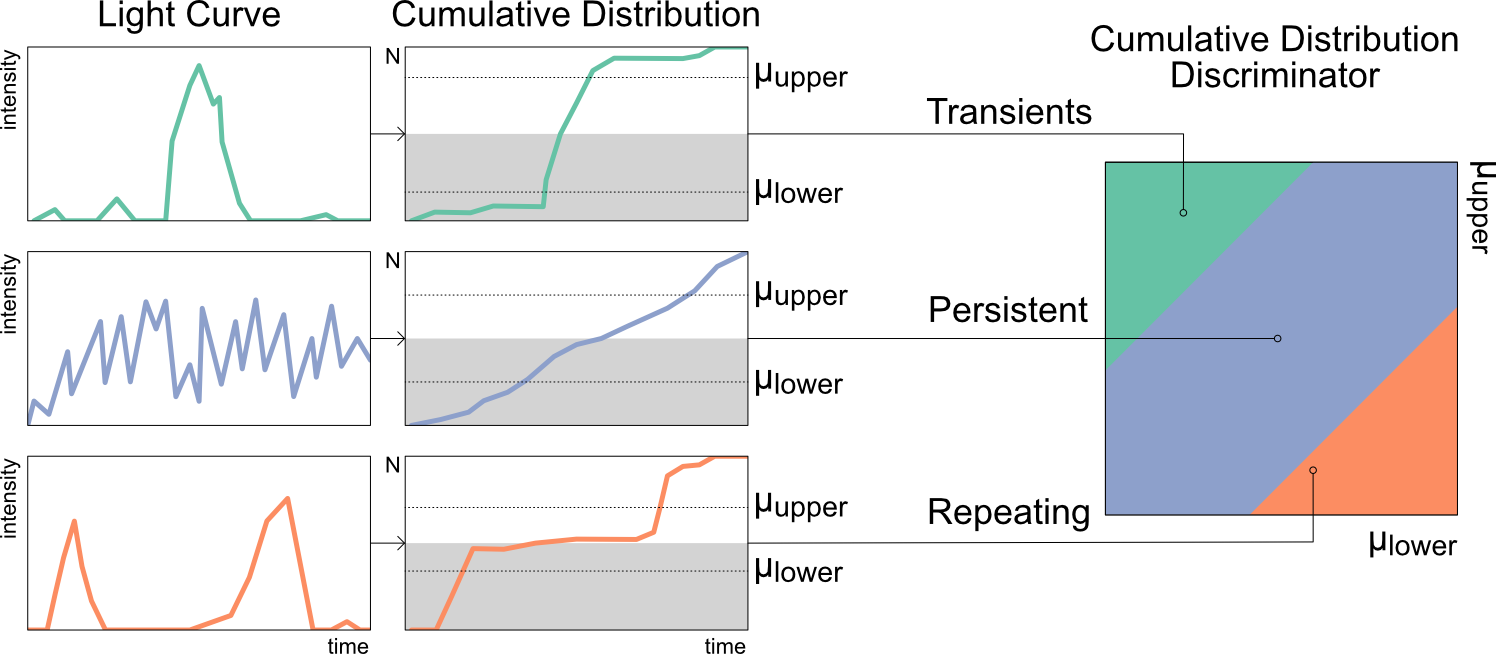}
    \centering
	 \caption{Source classification method using CuDiDi.
	Left: Sketches of representative light-curve behaviour, which illustrate typical variability patterns for persistent, repeating, and transient sources.
	Centre: Corresponding CDFs. The y-axis shows the cumulative number of counts $N$, normalised by the total so that each distribution reaches unity. Each CDF is split at the median (0.5) into lower (0–0.5, shaded light grey) and upper (0.5–1.0) halves. The mean values of each half, $\mu_{\text{lower}}$ and $\mu_{\text{upper}}$, are indicated.
	Right: CuDiDi plot showing $\mu_{\text{lower}}$ vs. $\mu_{\text{upper}}$ for each source. The diagonal band (±0.2 around $\mu_{\text{upper}}$ = $\mu_{\text{lower}}$ + 0.5) defines the persistent region. Sources above the band are classified as transients, while sources below the band are considered repeating sources. This visualisation shows how CuDiDi uses statistical asymmetries in cumulative flux distributions to separate source classes.
  }
	 \label{fig:cudidi}
\end{figure*}
The choice of the band width was guided empirically by visual inspection of representative examples, balancing sensitivity to genuine transient behaviour against robustness to statistical fluctuations and noise.
We also find that the selected threshold is somewhat stricter in terms of transient selection than previous searchers, but sufficiently low to recover the majority of the previously identified FXTs (see Sect.~\ref{sec:discussion}).  

To illustrate the effectiveness of the CuDiDi, we present in \autoref{fig:example_lcs} a set of representative sources. The top row shows three transient sources, while the bottom row displays three non-transient sources chosen from the same observations for comparison. The light curves (left column) highlight the contrasting temporal behaviours: transients exhibit sharp, localised bursts of activity, whereas non-transient sources show more uniform or recurrent variability. The corresponding cumulative distributions (middle column) emphasize this difference, with transients producing strongly asymmetric distributions compared to the smoother rise of persistent sources.
The right column summarises these properties in the CuDiDi plane, where each source is represented by its mean values $\mu_{\text{lower}}$ and $\mu_{\text{upper}}$ from the lower and upper halves of the cumulative distribution. The green shaded area indicates the region within which most transient sources fall. This example illustrates how the chosen band width provides a practical threshold for distinguishing source classes: narrower bands risk excluding borderline transients, whereas wider bands would allow contamination from persistent sources. 

The choice of time binning has a negligible impact on the values of $\mu_{\text{lower}}$ and $\mu_{\text{upper}}$ as long as the bins are small. For the six examples shown in Figure~3, the values change by $<10^{-4}$ if the time bins are increased from 1~s to 3.2~s (the frame time of the observations), which implies a negligible impact on the transient classification. By contrast, increasing the time bins to 1~ks for the same sources changes the values of $\mu_{\text{lower}}$ and $\mu_{\text{upper}}$ by up to 0.03, which could change the classification of sources near the boundaries (see also \autoref{app:bkg} for examples with 5~ks bins). 
The impact of increasing the time bins is expected to be larger for shorter observations and for sources with low count rates, which justifies our approach of using a uniform 1~s binning for all observations. Only sources that satisfied the CuDiDi criterion were retained for further analysis, leading to the exclusion of 306,136 additional sources, leaving 2,727 transient candidates.

\begin{figure*}[htbp]
	 \includegraphics[width=\textwidth]{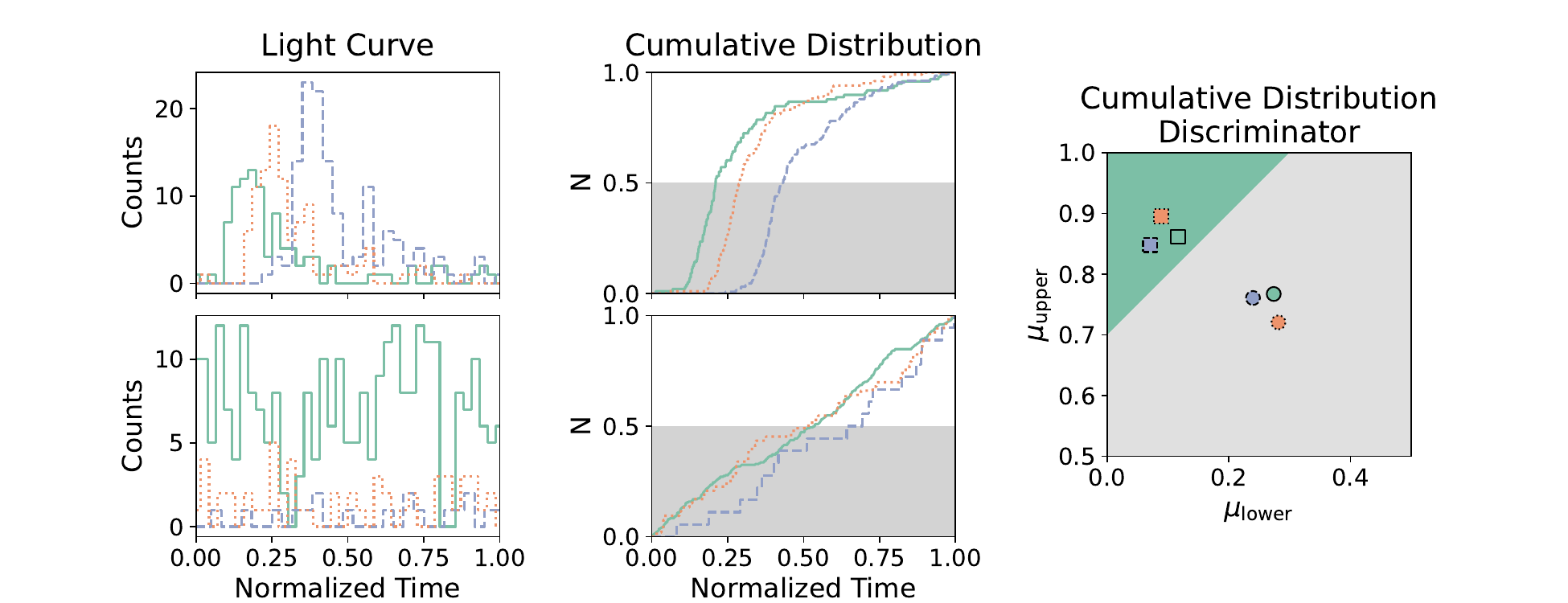}
    \centering
    \caption{Example X-ray light curves and their classification with CuDiDi.
Left column: Total counts binned at 1~ks for visual clarity and normalised in time for three representative sources per row. Transient sources are shown in the top row, and non-transient sources in the bottom row.
Middle column: Corresponding CDFs, obtained from the light curves with 1~s bins, normalised to unity. The shaded grey area indicates the lower half (0–0.5) of the cumulative fraction.
Right column: CuDiDi diagram showing $\mu_{\text{lower}}$ vs. $\mu_{\text{upper}}$, where $\mu_{\text{lower}}$ and $\mu_{\text{upper}}$ denote the mean values of the lower and upper halves of the cumulative distributions, respectively. Transient sources are marked as squares, and non-transient sources are shown as circles. The shaded green band indicates the transient region, defined as $+0.2$ above the diagonal $\mu_{\text{upper}} = \mu_{\text{lower}} + 0.5$. Sources in this region are classified as transients, while those below fall into the non-transient category. The colours and line styles in all panels indicate observation ID: solid green line = 7048, dashed blue line = 16174, dotted orange line = 20120. This demonstrates how CuDiDi captures the distinct variability patterns of transient and non-transient sources.}
	 \label{fig:example_lcs}
\end{figure*}

\subsubsection{(c) Quick-look analysis}  
Each candidate passing the automated criteria was subjected to a visual quick-look inspection of the \textit{Chandra} images. In this step, we checked for obvious instrumental artefacts, in particular the pronounced readout streaks that can mimic transient sources. Candidates affected by such artefacts were excluded.
This visual inspection led to the exclusion of 1,307 sources. Only sources that passed all three stages—cross-matching, CuDiDi filtering, and visual inspection—were classified as transient candidates and retained for detailed follow-up analysis.

Out of the 750,169 unique sources detected across all ACIS observations, 1420 passed the full sequence of selection criteria and were classified as transient candidates. The number or sources excluded at each stage of the analysis is summarised in \autoref{fig:flowchart}. 

\subsection{Diagnostic flags}
\label{sec:flags}

To facilitate further investigation and vetting of individual candidates, a series of diagnostic flags were added to the catalogue. These are summarised in \autoref{tab:diagnostic flags} and discussed below. This approach allows us to define a golden sample of high-confidence transients, while preserving the broader dataset for users interested in exploring more ambiguous or potentially interesting cases.

\begin{table*}
\caption{
Overview of diagnostic flags included in the transient candidate catalogue.}
\label{tab:diagnostic flags}
\centering
\begin{tabular}{llp{9.5cm}r}
	\hline\hline
	Flag Name & Type & Description & Count \\ 
	\hline
	\texttt{non\_unique} & Boolean & True if the transient is matched with a source detected at low significance in other \textit{Chandra} observations (sources detected at $>3\sigma$ in more than one observation are already excluded). & 58 \\
	\texttt{snr\_match} & Boolean & True if the source is associated with a known Galactic SNR (e.g. Cas~A, Tycho), based on observation ID cross-matching. & 5 \\
	\texttt{extended\_fov} & Boolean & True if the source lies in or near an extended Galactic structure identified by manual inspection of the FOV. & 13 \\
	\texttt{readout\_streak} & Boolean & True if the source overlaps with a known instrumental readout streak, commonly from bright sources like Cyg~X$-$1. & 29 \\
	\texttt{edge\_affected} & Boolean & True if the source region (as determined by \texttt{wavdetect}) is not fully covered due to detector dithering; computed via \texttt{dither\_region}. & 392 \\
	\texttt{bkg\_correlated} & Boolean & True if the Pearson correlation between the source and background light curves during the $T_{90}$ interval exceeds 0.15, indicating likely background flaring. & 49 \\
	\texttt{start\_or\_end} & Boolean & True if the $T_5$ or $T_{95}$ time falls within the first or last 5\% of the observation, indicating the event may lie at the edge of the observing window. & 266 \\
	\hline
\end{tabular}
\tablefoot{These flags helped us to identify likely artefacts or ambiguous cases that may not represent genuine transient events. Out of a total of 1420 candidates, 765 form a golden sample with none of the listed flags set to true.
}
\end{table*}

\paragraph{Uniqueness flag (\texttt{non\_unique})} 
While sources detected at high significance in more than one observation were excluded in the cross-matching described above, we performed an additional check by matching the transient candidates against all other sources detected by \texttt{wavdetect} using a radius of  3\arcsec\. This finds matches with sources detected at a significance below $3\sigma$, which may be spurious or due to weak nearby sources, but could also indicate that the transient is in fact a recurring source. Candidates with at least one positional match in a different observation from our database were flagged as non-unique.

\paragraph{Galactic exclusion flags (\texttt{snr\_match} and \texttt{extended\_fov})}
Since our primary goal is to identify a high-purity sample of transient candidates, we excluded sources associated with prominent extended X-ray sources, which add uncertainties to the source detection and analysis. Specifically, we flagged all candidates found in observations targeting well-known supernova remnants (SNRs) such as Cas A and Tycho, identified through observation ID matching, under the flag \texttt{snr\_match}. Additionally, extended Galactic sources were flagged with \texttt{extended\_fov} via manual inspection of FOV images.

These associations do not definitively rule out a background transient event located behind the foreground structure, so we retained the sources in the full candidate catalogue but flagged them for caution. We stress, however, that the vast majority of these transients are expected to be spurious. Based on the surface area of the main SNRs and the number of observations, we expect less than one genuine transient in our sample to be superposed on them by chance.

\paragraph{Artefact flags (\texttt{readout\_streak} and \texttt{edge\_affected})}
Two main observational artefacts were identified that could mimic transient behaviour. First, sources lying along readout streaks\footnote{\url{https://cxc.cfa.harvard.edu/ciao/dictionary/readout_streak.html}} \citep{2005ASPC..347..478M, 2010ApJS..189...37E}, particularly in observations of bright sources such as Cyg X$-$1, were flagged following visual inspection of the images, using the flag \texttt{readout\_streak}. Although the quick-look analysis excluded the most obvious cases, these flags capture more ambiguous sources that were not removed in that step.

Secondly, sources near the detector edge may show apparent variability due to dithering motion\footnote{\url{https://cxc.cfa.harvard.edu/ciao/why/dither.html}} moving the source region partially on and off the detector \citep{2010ApJS..189...37E, 2011ApJS..194...37P}. To mitigate this, we used the CIAO tool \texttt{dither\_region}\footnote{\url{https://cxc.cfa.harvard.edu/ciao/ahelp/dither_region.html}} to estimate the fractional exposure of the source region. Any source with a less than full coverage (i.e. a coverage fraction < 1.0) was flagged as \texttt{edge\_affected}.

\paragraph{Background flag (\texttt{bkg\_correlated})}
Since the CuDiDi analysis is based on total (non-background-subtracted) counts, background flaring can produce artificial variability in the light curve. To identify such cases, we computed the Pearson correlation coefficient between the source and background light curves, restricted to the interval between the 5th and 95th percentiles of the source photon arrival times (i.e. the $T_{90}$ interval). Sources with a correlation coefficient exceeding 0.15 were flagged as (\texttt{bkg\_correlated}) and are likely dominated by background flaring. This threshold was empirically determined by testing several candidate values and visually inspecting sources around the corresponding decision boundary to assess how the flagging behaviour changed.

Although standard GTI filtering excludes intervals with high global background, which typically removes strong, long-duration flares, it does not always capture low-level or localised background variability. In our sample, 49 sources showed significant correlation between source and background light curves, indicating that some background fluctuations remained even after GTI filtering. This additional check helps to identify and exclude false positives that would otherwise contaminate the transient candidate list.

\paragraph{Start- or end-of-observation flag (\texttt{start\_or\_end})}
Transient classification requires sufficient temporal baseline to establish quiescence before and after the event. Sources whose variability occurred at the very beginning or end of the observation are ambiguous in this regard. We therefore flagged any source for which the start ($T_{5}$) or end ($T_{95}$) of the light curve overlapped with the first or last 5\% of the observation duration.

Based on the full set of diagnostic flags described above, we define a high-confidence golden sample of transient candidates as those for which all cautionary flags are set to \texttt{False}. This conservative subset is intended to minimise contamination from instrumental artefacts, background variability, and foreground Galactic sources.
In Sect.~\ref{sec:properties}, we explore the spatial distribution, temporal characteristics, and X-ray spectral properties of the 765 sources in this golden sample.

\section{X-ray properties}\label{sec:properties}

\subsection{General properties}
\autoref{fig:exposure-times} shows the distribution of exposure times for all 20,212 \textit{Chandra} ACIS observations included in our search, along with the subset containing the 765 transient candidates. The exposure times span a wide range, from brief observations of less than 1~ks to deep observations exceeding 100~ks. The overall distribution peaks around 30~ks, with a mean of 23~ks and a median of 16~ks, consistent with the typical duration of \textit{Chandra} imaging observations \citep{2011ApJS..194...37P}.

Importantly, transient candidates (hatched bars) are found across the full range of exposure times. Their distribution closely tracks that of the full dataset, indicating that the detection method does not significantly favour either short or long observations. In particular, the presence of candidates in short ($\lesssim$5~ks) and long ($\gtrsim$50~ks) exposures demonstrates the method’s robustness across varying observational depths. This also suggests that the transient events we identify are not systematically tied to any particular exposure regime and that detection efficiency is not strongly biased by exposure time.

\begin{figure}[htbp]
	 \resizebox{\hsize}{!}{\includegraphics[width=.9\textwidth]{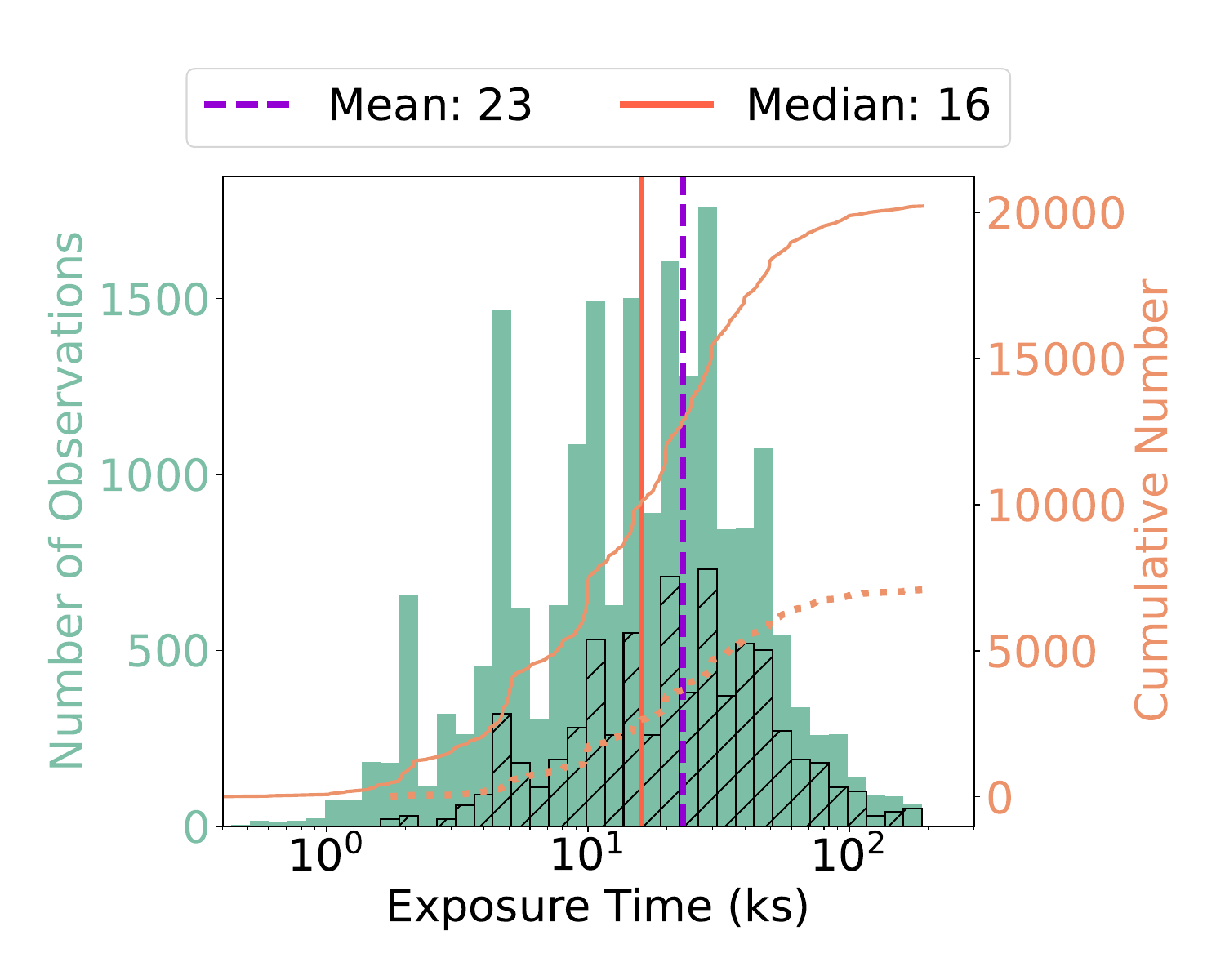}}
      \centering
	 \caption{Histogram (green; left y-axis) and cumulative distribution (orange lines; right y-axis) of the exposure times for the 20,212 \textit{Chandra} observations. The vertical dashed purple and solid dark orange lines indicate the mean and median exposure times of the full sample, respectively. The distribution for 765 transient candidates  is scaled by a factor of 10 for visual clarity (hatched bars and dotted orange line). 
  }
	 \label{fig:exposure-times}
\end{figure}

To illustrate the identification of transients with the CuDiDi method, we applied it to all 750,169 light curves in our sample, with the resulting distribution shown in \autoref{fig:cudidi-heatmap}. The majority of sources form a tight cluster centred near $(\mu_{\text{lower}},\mu_{\text{upper}})\approx(0.25,0.75)$, consistent with constant or weakly variable emission. The figure also shows the boundaries used for defining transient and persistent sources.

The 765 sources that constitute our golden sample are overplotted in \autoref{fig:cudidi-heatmap} as light-blue points. Their concentration just above the upper edge of the band reflects the applied threshold rather than an intrinsic segregation of the population. Within the transient region, these sources span a wide range of asymmetries in their cumulative distributions, indicating diverse variability behaviours rather than a single characteristic profile. This spread suggests that CuDiDi captures a continuum of transient morphologies, from brief, 
sharply peaked events to more extended episodes of enhanced emission. The empirical cut therefore provides a practical, though somewhat arbitrary, means of separating transients from the dominant non-transient population. In contrast, relatively few sources are found below the persistent band. While this region could in principle contain sources with repeating or periodic flaring behaviour within a single observation, our classification boundaries were not tuned or validated for such cases, and we do not further explore this population in this work.

\begin{figure}[htbp]
	 \resizebox{\hsize}{!}{\includegraphics[width=.9\textwidth]{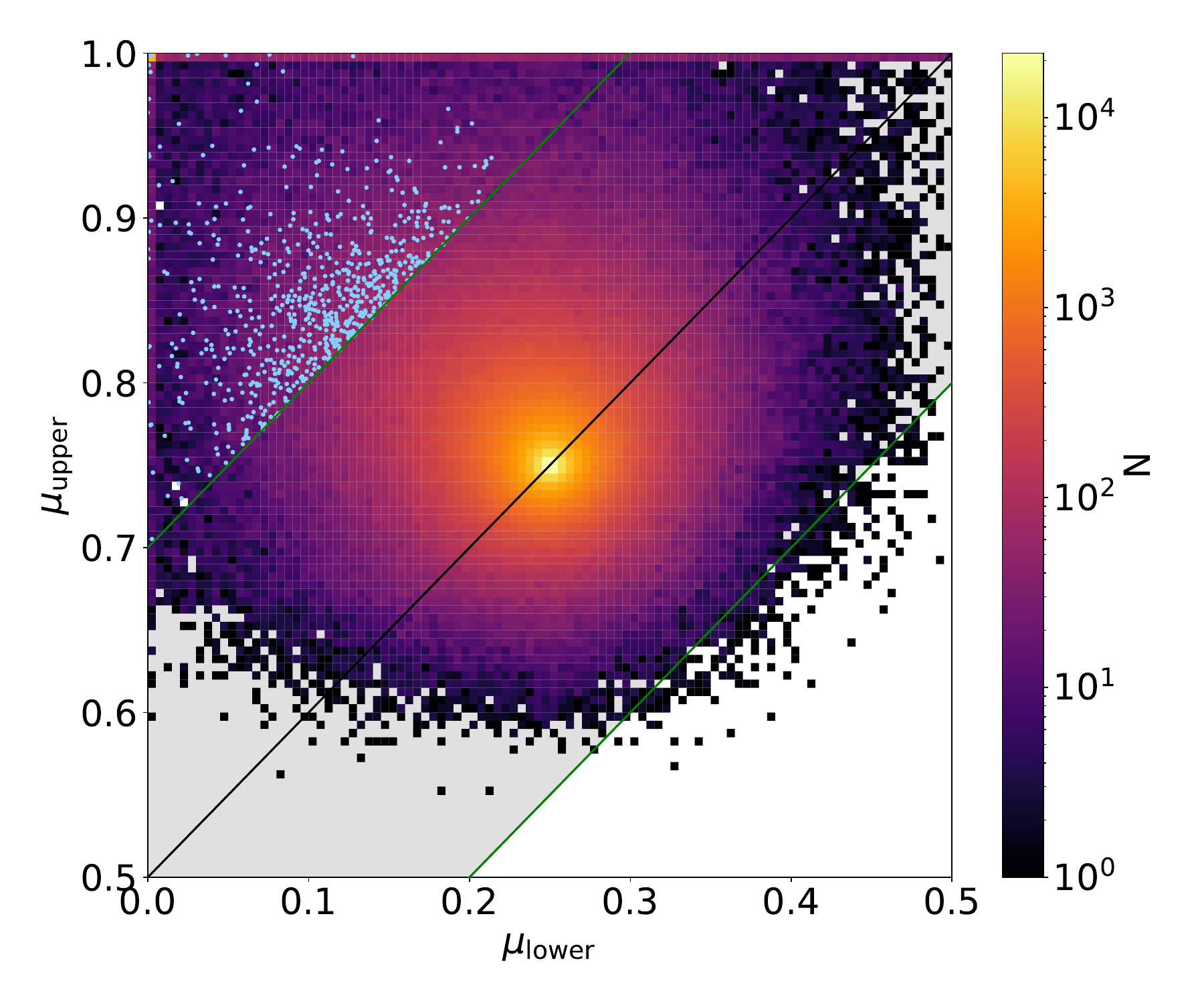}}
      \centering
	 \caption{CuDiDi distribution of the photon-arrival-time asymmetry, showing  $\mu_{\text{lower}}$ vs. $\mu_{\text{upper}}$ for all sources we analysed (background-density map). The diagonal black line indicates the locus of temporally symmetric light curves, $\mu_{\text{upper}}=\mu_{\text{lower}}+0.5$. The green lines mark the empirical persistent band, defined as $|\,\mu_{\text{upper}}-(\mu_{\text{lower}}+0.5)\,|\leq0.2$. Sources within and below this band are considered non-transient, while sources above are classified as transients. The 765 transient candidates in our golden sample are overplotted as light blue points. By construction they lie above the band, but their spread within this region reflects a wide range of asymmetry strengths. The bulk of the overall population clusters around $(0.25,0.75)$, as expected for approximately constant emission.}
	 \label{fig:cudidi-heatmap}
\end{figure}

We next examined the brightness distribution of the transient candidates, based on the net background subtracted counts. \autoref{fig:totnetcts} shows a histogram spanning from 1 to over 200 counts, with a strong peak at the low-count end. The median count is 9, and the mean is 12. An inset zooms into the 0--50 count range, where the bulk of the population lies.
The large number of low-count candidates reflects the method’s sensitivity to faint or short-duration transients, which may not stand out in time-averaged images but exhibit clear temporal structure.
At the same time it is clear that sources with a small number of counts are more likely to belong to the category of transients that exhibit occasional flares in addition to persistent faint emission, where the persistent emission would be below the \textit{Chandra} sensitivity limit. Our check for recurring sources would, however, have excluded any such sources that have been observed to flare more than once. 

\begin{figure}[htbp]
	 \resizebox{\hsize}{!}{\includegraphics[width=.9\textwidth]{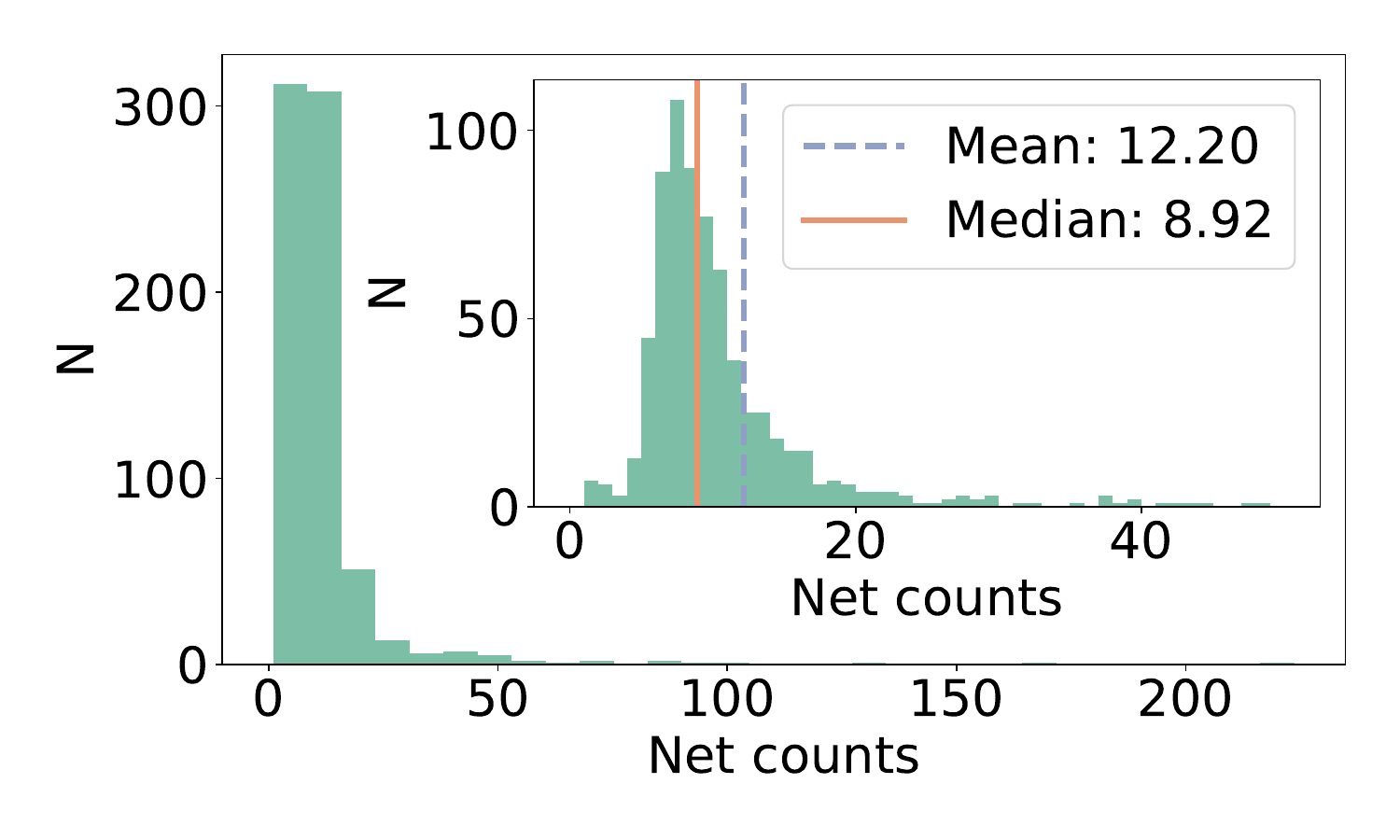}}
      \centering
	 \caption{Distribution of net counts for all 765 transient candidates. The histogram spans a wide range, with a mean of 12 and a median of 9 counts per candidate. The inset shows a zoomed-in view of the 0–50 count range to highlight the peak and shape of the low-count population. The prevalence of low-count events reflects the sensitivity of the method to faint or short-duration transients.}
	 \label{fig:totnetcts}
\end{figure}

\subsection{Spatial distribution}
If the transient candidates are predominantly extragalactic in origin, their sky distribution should be approximately isotropic, aside from modulations due to spatial variations in the total exposure time of all the \textit{Chandra} observations.
 \autoref{fig:sky-image} shows the distribution of all 765 candidates in Galactic coordinates.
This shows that the transients are distributed across the whole sky, without any strong large-scale anisotropy. 
In particular, there is no strong concentration of candidates toward the Galactic plane, which would be expected if a substantial fraction were tied to Galactic stellar populations or compact objects. This suggests that many of the candidates may be extragalactic in nature, although definitive classification requires further multi-wavelength analysis.

\begin{figure}[htbp]
	 \resizebox{\hsize}{!}{\includegraphics[width=.9\textwidth]{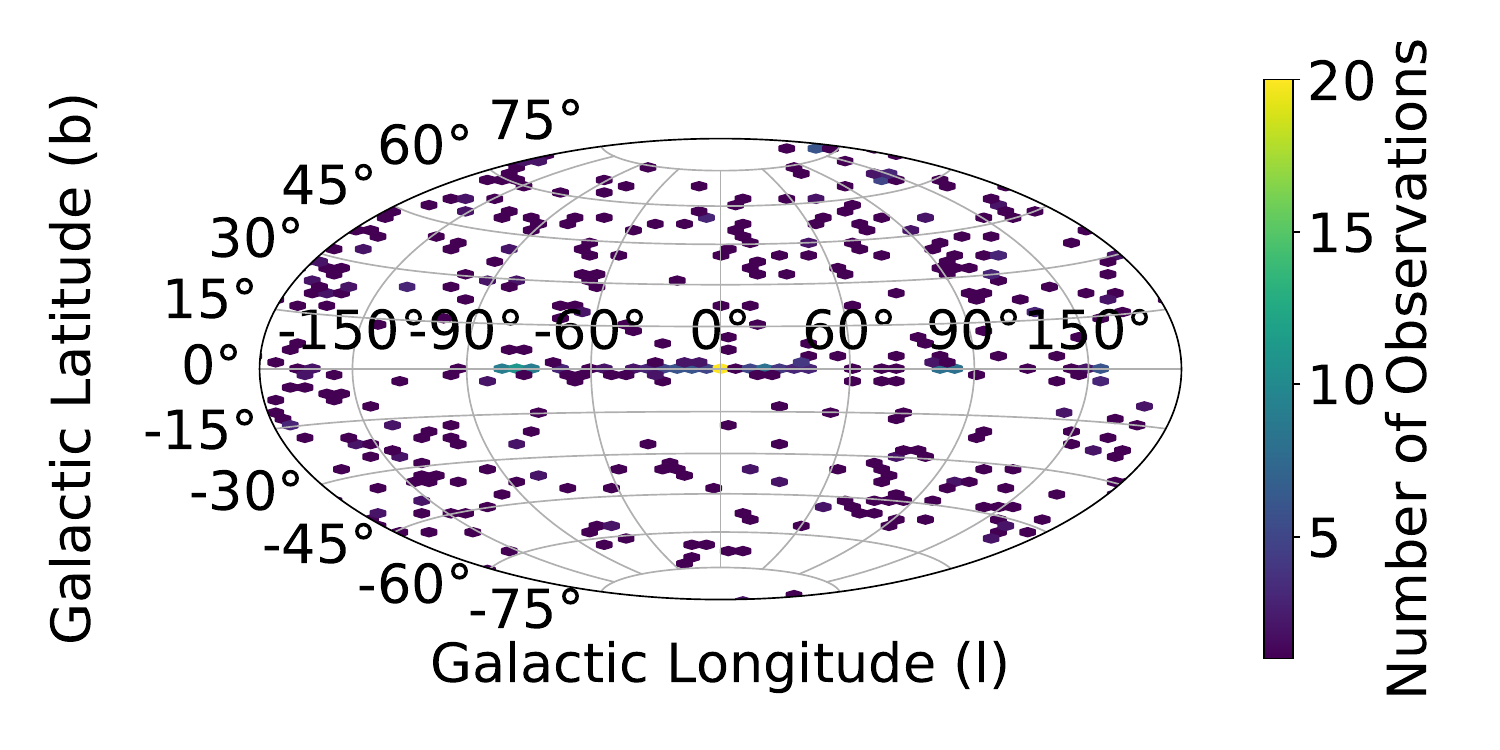}}
      \centering
	 \caption{Sky distribution of the 765 transient candidates in Galactic coordinates using a Hammer-Aitoff projection. The hexbin density map shows the number of candidates per bin across the sky.}
	 \label{fig:sky-image}
\end{figure}

\subsection{Temporal properties}\label{sec:temporal properties}
To quantify the characteristic duration of each transient, we measured the \(T_{90}\) interval, defined as the time over which 90\% of the total source counts are accumulated, between the epochs at which 5\% (\(T_{5}\)) and 95\% (\(T_{95}\)) of the counts are reached. A direct determination from the raw cumulative distributions was found to be unreliable in some cases due to statistical fluctuations and irregular light-curve shapes. We therefore fitted the cumulative total counts of each candidate with a sigmoid function of the form
\[
f(t) = \frac{L}{1 + \exp\!\left[-k \,(t - t_{0})\right]},
\]
where \(L\) is the asymptotic total counts, \(t_{0}\) is the midpoint, and \(k\) is the steepness parameter. The fit provides a smooth monotonic representation of the cumulative distribution, from which \(T_{5}\) and \(T_{95}\) were determined as the times at which the fitted curve reaches 5\% and 95\% of \(L\). The duration is then given by \(T_{90} = T_{95} - T_{5}\).
\autoref{fig:t90} shows the resulting distribution, which spans over five orders of magnitude, from second durations to a substantial fraction of the longest \textit{Chandra} observations. The distribution has a median of approximately 10{,}000~s and a mean of 14{,}000~s, indicating that most events are relatively long-lived but with substantial spread.

The inset zooms into the short-duration regime, revealing a substantial tail of fast events with \(T_{90} < 60\)~s. Notably, four candidates exhibit a measured \(T_{90} = 0\)~s. These extreme cases 
are discussed in more detail in Sect.~\ref{sec:discussion-0s}, where we find that they are most likely genuine transients, while the \(T_{90} = 0\) is an artefact of the sigmoid fit. 
The broad distribution of \(T_{90}\) values reflects the sensitivity of the method to rapid and slowly evolving events.

\begin{figure}[htbp]
	 \resizebox{\hsize}{!}{\includegraphics[width=.9\textwidth]{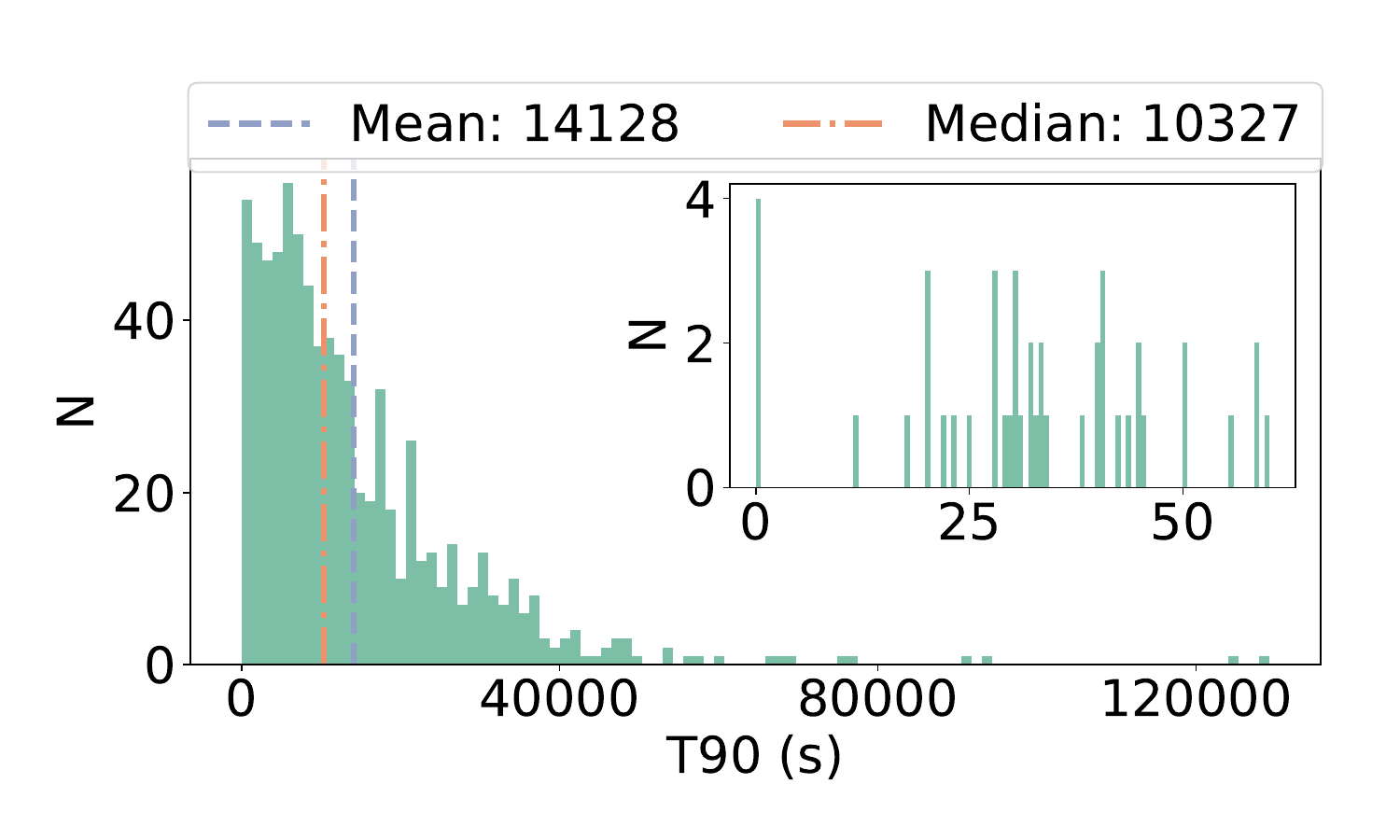}}
      \centering
	 \caption{Distribution of \(T_{90}\) durations (the time interval containing 90\% of the total counts) for all 765 transient candidates. The zoomed inset highlights the short-duration regime (\(T_{90} \leq 60\) s). The durations span from effectively zero to over 120,000~s, with a mean of approximately 14,000~s and a median of 10,000~s. Four candidates with \(T_{90} = 0\) s are discussed in Sect.~\ref{sec:discussion-0s}.}
	 \label{fig:t90}
\end{figure}

To assess the duration of each transient in the context of its observation, we compute the fractional timescale \(T_{90} / T_{\text{obs}}\), where \(T_{\text{obs}}\) is the total exposure time. \autoref{fig:t90obsfrac} shows the distribution of this quantity across all 765 candidates.
A prominent peak near zero reflects the population of short-duration events. The rest of the distribution is relatively broad, extending to \(T_{90} / T_{\text{obs}} \sim 0.8\), indicating a significant subset of events that last throughout most of the exposure. This diversity in relative timescales suggests that the candidate sample includes multiple transient classes with differing physical origins.

\begin{figure}[htbp]
	 \resizebox{\hsize}{!}{\includegraphics[width=.9\textwidth]{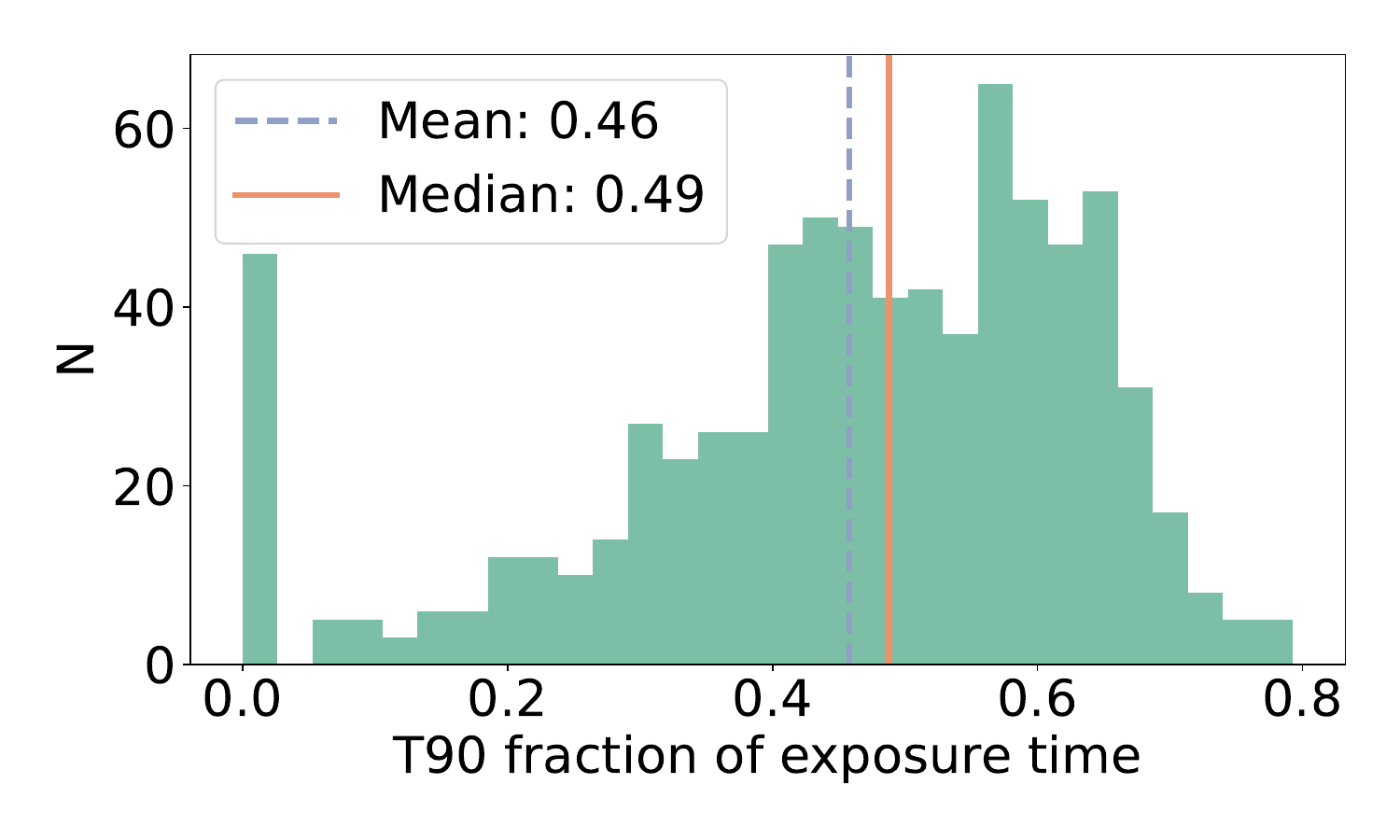}}
      \centering
	 \caption{Distribution of \(T_{90} / T_{\text{obs}}\), the fractional duration of each transient relative to its parent observation. The sharp peak near zero reflects a population of very short transients, while the broader tail extending up to \(\sim 0.8\) indicates events that persist for a substantial fraction of the exposure.}
	 \label{fig:t90obsfrac}
\end{figure}

\subsection{Spectral properties}  

To characterise the spectral properties of our transient candidates, we calculated the hardness ratio (HR), a commonly used metric for low-count X-ray sources. The HR is defined as  
\[ 
    \text{HR} = \frac{H - S}{H + S},  
\] 
where \(H\) and \(S\) correspond to the number of photons detected in the hard (2--7~keV) and soft (0.5--2~keV) bands, respectively.  
We obtained valid HR measurements for 424 candidates. The remaining sources lack HR values due to empty background regions or errors encountered during spectral extraction, which prevented a reliable separation of counts into soft and hard bands.  
\autoref{fig:hr} shows the distribution of hardness ratios for the 424 candidates. The clear peak toward negative HR values indicates that the majority of sources are dominated by soft X-ray emission, as expected since the source detection was performed on images in the soft band.
We also investigated potential trends between HR and event duration (\(T_{90}\)), but found no significant correlations or evidence of clustering in the distribution. This is shown in \autoref{fig:t90hr}.

\begin{figure}[htbp]  
	 \resizebox{\hsize}{!}{\includegraphics[width=.9\textwidth]{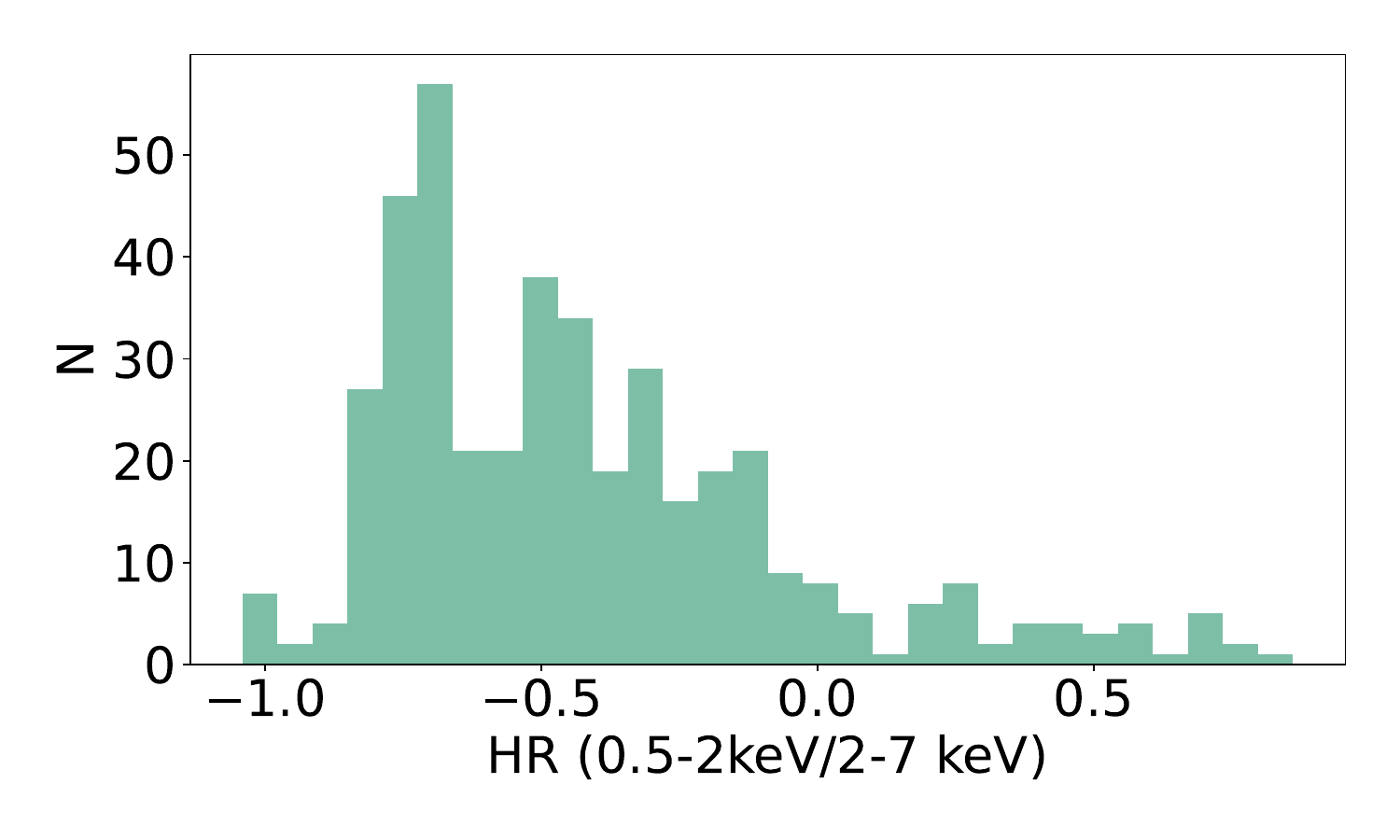}}  
      \centering  
	 \caption{Distribution of HRs for the transient candidates. The HR is defined as \(\mathrm{HR} = (H - S)/(H + S)\), where \(H\) and \(S\) are counts in the hard (2--7~keV) and soft (0.5--2~keV) bands, respectively. The distribution peaks toward negative values, indicating that the majority of candidates exhibit relatively soft X-ray spectra.}  
	 \label{fig:hr}  
\end{figure}  
\begin{figure}[htbp]  
	 \resizebox{\hsize}{!}{\includegraphics[width=.9\textwidth]{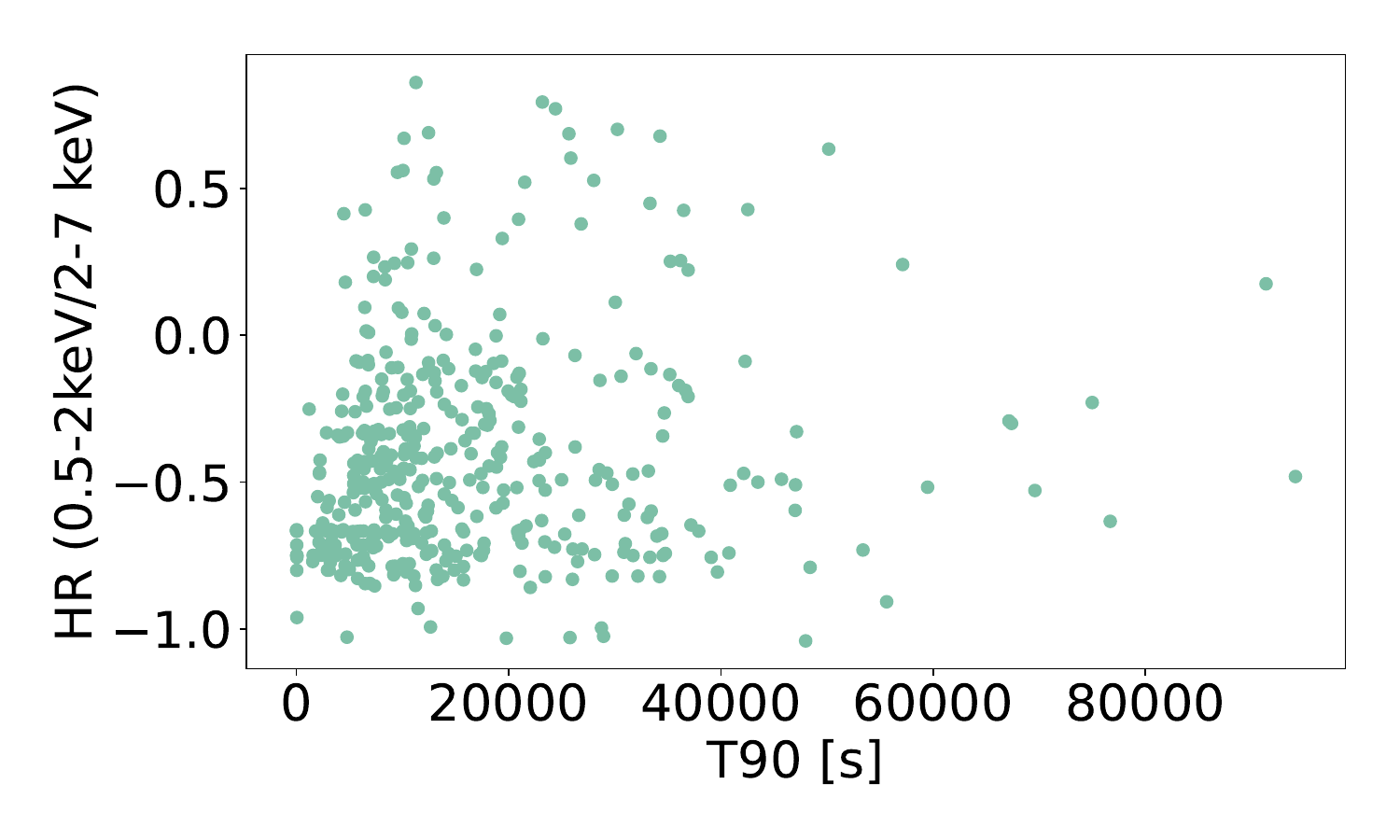}}  
      \centering  
	 \caption{Distribution of the HR as a function of \(T_{90}\) for the transient candidates.}  
	 \label{fig:t90hr}  
\end{figure}

\section{Discussion}\label{sec:discussion}

\subsection{The T90 = 0s population}
\label{sec:discussion-0s}
Four candidates stand out in the duration distribution (\autoref{fig:t90}, inset), where they appear formally consistent with $T_{90} = 0$~s. This behaviour arises from the way durations are measured rather than from physical properties. Specifically, $T_{90}$ is derived from a sigmoid fit to the cumulative photon arrival distribution (Sect.~\ref{sec:temporal properties}), with $T_{5}$ and $T_{95}$ defined as the epochs when the fitted curve reaches 5\% and 95\% of the total counts. For short transients embedded within long observations, the fit becomes extremely steep, and the resulting $T_{90}$ values collapse toward zero, even though the events span multiple seconds in real time.
The light curves of these candidates (\autoref{fig:t90z_lc}) illustrate this point. In the zoomed-in insets, the photon arrival times span intervals from $\sim$1~s to $\sim$32~s. In the most extreme case, X-ray transient (XT)~160625, several photons are recorded within what appears as a 1~s interval, followed by additional isolated photons before and after the peak, suggesting that the true activity may extend over a longer period. We note, however, that the instrumental frame time in these four observations is 3.14~s. To enable consistent comparison across the dataset, all light curves were extracted with a uniform 1~s binning, which oversamples the intrinsic resolution for many observations but preserves the finest accessible temporal structure. This choice does not bias the transient selection (see Sect.~\ref{sec:selection}) and allows flexible rebinning to coarser resolutions if desired.

\begin{figure*}[htbp]
	 \resizebox{\hsize}{!}{\includegraphics[width=.9\textwidth]{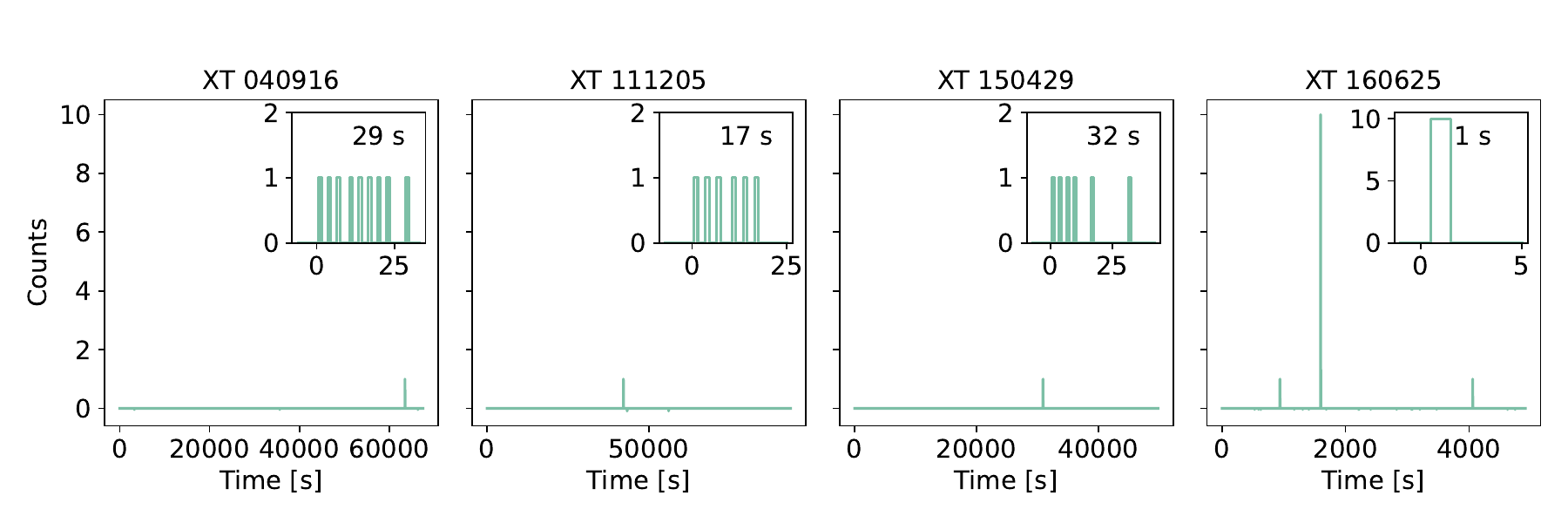}}
      \centering
	 \caption{Background-subtracted light curves of the transients with zero-duration $T_{90}$. The main panels show the full observation light curves, with the x-axes set such that 0~s corresponds to the start of the observation. The insets zoom in on the transient events, with x-axes shifted so that the first bin with signal corresponds to 0~s, and the durations of the events are indicated. Counts are binned to 1~s. These events last between $\sim$1 and $\sim$32~s.
     }
	 \label{fig:t90z_lc}
\end{figure*}

The basic observational properties of the four candidates are listed in \autoref{tab:t90eq0_population}. Two of the events (XT~040916 and XT~160625) were detected under conditions that limit duration constraints — one near the end of a 67.5~ks observation, and one during a short 4.9~ks pointing. The other two (XT~111205 and XT~150429) were observed in the middle of long exposures (93.9~ks and 49.8~ks, respectively), making it unlikely that extended flaring activity was missed.
\begin{table*}
\centering
\caption{Selected candidates with zero-duration $T_{90}$.}
\label{tab:t90eq0_population}
\begin{tabular}{lcccccl}
\hline\hline
Source ID & Obs. ID & RA & Dec & Obs. Time & Exp. Time & Notes \\
& & (deg) & (deg) & YYYY Mmm dd hh:mm:ss & (ks) & \\
\hline
XT 040916 & 5024 & 161.161145 & 58.733960 & 2004 Sep 16 06:54:50 & 67.5 & close to end of observation\\
XT 111205 & 13610 & 100.216376 & 9.538021 & 2011 Dec 05 01:53:23 & 93.9 & NGC 2264\\
XT 150429 & 17632 & 206.620123 & 26.423577 & 2015 Apr 29 09:07:16 & 49.8 & \\
XT 160625 & 18131 & 56.775122 & -36.777104 & 2016 Jun 25 01:10:04 & 4.9 & \\
\hline
\end{tabular}
\tablefoot{Each row includes the source ID, observation ID, sky position, observation start time, approximate exposure time and relevant notes.}
\end{table*}
In summary, the apparent $T_{90} = 0$~s values should not be interpreted as single-frame transients. Instead, they reflect limitations of the sigmoid-based estimator when applied to very short events in long exposures. Improved methods, for example by incorporating direct photon arrival distributions, may yield more accurate duration measurements in the future.
Nevertheless, these four transients still probe the extreme end of the timescale distribution for the transients identified in our search, and we therefore examine their properties in more detail below.

Before considering physical interpretations, we checked for potential instrumental origins. The four candidates are not concentrated near chip edges, nor are they located in regions with known extended sources (e.g. the vicinity of Cas A). Events of that kind would have been flagged by our diagnostic criteria and would not be included in the high-confidence (golden) sample.
None of the candidates were flagged by the standard \textit{Chandra} artefact diagnostics (e.g. streaks, hot pixels, cosmic-ray afterglows), which are designed to identify spurious detections of instrumental origin. We therefore consider it unlikely that these events arise from known detector effects.
With instrumental origins unlikely, we can turn to the spectral properties of these events.

The spectra of the four transients were fitted with XSPEC version 12.14.1 \citep{Arnaud1996}, using the C-stat fit statistic \citep{Cash1979}. The spectra were grouped to one count per bin. Parameter uncertainties were obtained with the \texttt{error} command and correspond to 90$\%$ confidence intervals for one parameter of interest. 
The spectra are reasonably well fitted by blackbody models or soft power laws, as summarised in  \autoref{tab:t90z_fitpar}. The only exception is XT~160625, for which the parameters of the power-law model could not be constrained. The Galactic column density of hydrogen ($N_{\rm H}$) in the direction of these transients is $7.61 \times 10^{19}\ \rm cm^{-2}$ (XT~040916), $4.43 \times 10^{21}\ \rm cm^{-2}$ (XT~111205), $1.00 \times 10^{20}\ \rm cm^{-2}$ (XT~150429), and $1.75 \times 10^{20}\ \rm cm^{-2}$ (XT~160625), determined from the NHtot tool\footnote{\url{https://www.swift.ac.uk/analysis/nhtot/}} \citep{Willingale2013}. These levels of absorption have a negligible impact on the fit results in all cases except XT~111205, for which we also report results that include absorption fixed at the Galactic value in \autoref{tab:t90z_fitpar}, modelled with the XSPEC \texttt{tbabs} model \citep{Wilms2000}. The spectra of the four transients fitted with the blackbody model without absorption are shown in \autoref{fig:t90z_spec}.

The results in \autoref{tab:t90z_fitpar} highlight the very soft nature of the spectra, with $kT$ in the range $\sim$ 0.06--0.4~keV, or photon indices in the range $\Gamma \sim $ 2.5--7. 
This is also seen in \autoref{fig:t90z_spec}, which shows that XT~040916 is the only source with detected counts  above 2~keV. For reference, we report fluxes from the best-fit models in the full \textit{Chandra} energy range 0.5--7~keV in \autoref{tab:t90z_fitpar}. XT~0401916, XT~111205, and XT~150429 have similar unabsorbed fluxes in the range  $\sim 2-5 \times 10^{-15}\  {\rm erg\ s^{-1}\ cm^{-2}}$, while XT~160625 is brighter at  $\sim 7 \times 10^{-15}\  {\rm erg\ s^{-1}\ cm^{-2}}$.

\begin{figure*}[htbp]
	 \resizebox{\hsize}{!}{\includegraphics[width=.9\textwidth]{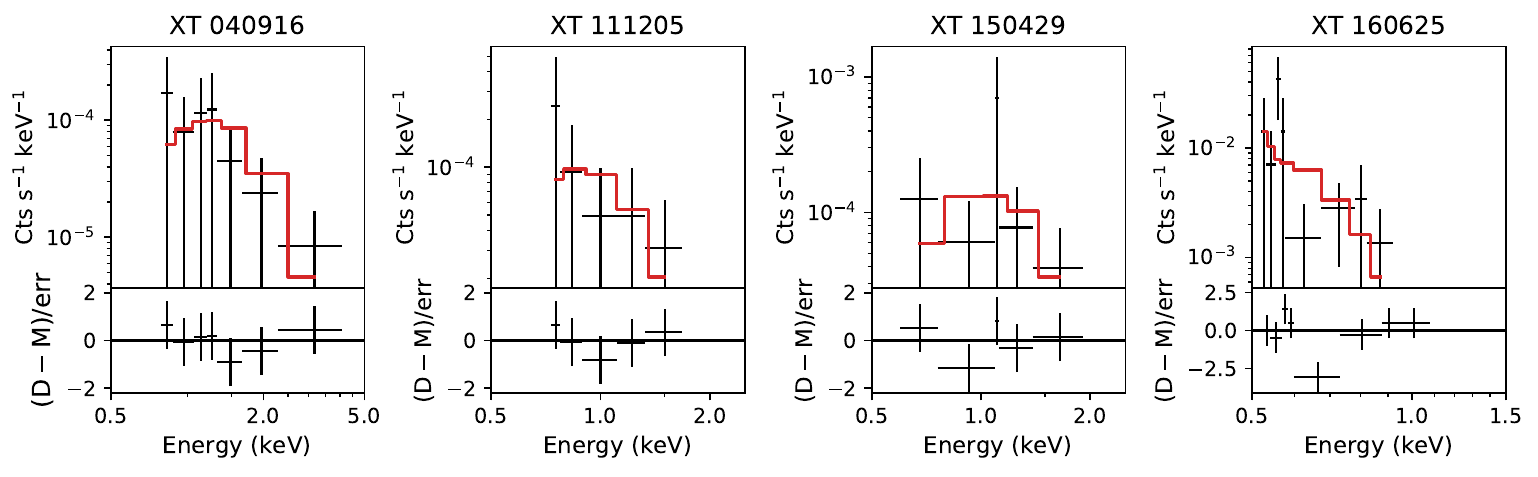}}
      \centering
	 \caption{Spectra of the transients with zero-duration $T_{90}$. The upper panels show the spectra (black crosses) and the best-fit blackbody models (red lines), while the lower panels show the residuals between the data (D) and models (M) normalised by the uncertainty. Note the different upper limits on the x-axes, set to cover the energy ranges where there are counts in the spectra.}
	 \label{fig:t90z_spec}
\end{figure*}

\begin{table*}
\centering
\caption{Results of spectral fits for candidates with zero-duration $T_{90}$.}
\label{tab:t90z_fitpar}
\begin{tabular}{lcccccc}
\hline\hline
Source ID & \multicolumn{3}{c}{Blackbody model} & \multicolumn{3}{c}{Power law model} \\ 
& $T$ & $F_{\rm 0.5-7\ keV}$ & C-stat/d.o.f   &$\Gamma$ & $F_{\rm 0.5-7\ keV}$ & C-stat/d.o.f  \\ \smallskip
& (keV) & (${\rm erg\ s^{-1}\ cm^{-2}}$) & & & (${\rm erg\ s^{-1}\ cm^{-2}}$)  & \\  \hline \smallskip
XT 040916 & $0.40^{+0.38}_{-0.16}$ & $2.5^{+2.5}_{-1.3} \times 10^{-15}$  & 1.84/7  & $2.5^{+1.9}_{-1.6}$ & $3.7^{+4.3}_{-1.9} \times 10^{-15}$ & 0.69/7 \\ \smallskip
XT 111205 & $0.17^{+0.27}_{-0.07}$ &  $1.9^{+3.0}_{-1.2} \times 10^{-15}$ &   1.39/4  & $4.7^{+3.9}_{-3.6}$ & $3.2^{+13.9}_{-2.1} \times 10^{-15}$ & 0.87/4 \\ \smallskip
 & $0.12^{+0.11}_{-0.04}$ & $1.0^{+1.4}_{-0.8} \times 10^{-14}$ & 2.07/4 & $7.0^{+3.0}_{-3.8}$ &  $2.9^{+14.4}_{-2.5} \times 10^{-14}$ & 1.25/4 \\ \smallskip
XT 150429 & $0.19^{+0.24}_{-0.07}$ &  $4.7^{+7.4}_{-2.9} \times 10^{-15}$ &   3.06/3  & $4.0^{+2.8}_{-2.9}$ & $6.9^{+9.8}_{-4.2} \times 10^{-15}$ & 2.8/3\\ \smallskip
XT 160625 & $0.062^{+0.021}_{-0.015}$  &  $7.3^{+5.5}_{-3.5} \times 10^{-13}$  &  9.84/8  & -- & -- & -- \\ 
\hline
\end{tabular}
\tablefoot{The photon index ($\Gamma$) of the power-law model could not be constrained for XT~160625. The second row for XT~111205 includes Galactic absorption fixed at $N_{\rm H} = 4.43 \times 10^{21} \rm cm^{-2}$, and the corresponding fluxes are unabsorbed. All other results are for models without Galactic absorption. The $N_{\rm H}$ of the other sources are in the range 0.76--1.75$ \times 10^{20}$ and have a negligible effect on the fits.
}
\end{table*}

We carried out a search for host galaxies and possible Galactic counterparts in order to gain further insight into the nature of the sources, and ideally be able to determine their luminosities. 
The aim was to assess whether these events could be plausibly associated with galaxies in their vicinity.
To this end, visual inspections were performed using multi-wavelength imaging data. Sky views centred on each candidate were generated with the Aladin Lite \citep{2022ASPC..532....7B} interface across several optical, infrared, and ultraviolet surveys, including the Sloan digital sky survey \cite[SDSS;][]{2012ApJS..203...21A}, Pan-STARRS \citep{2016arXiv161205560C}, the two micron all sky survey \cite[2MASS;][]{2006AJ....131.1163S}, the wide-field infrared survey explorer \cite[WISE;][]{2010AJ....140.1868W}, the digitized sky survey \cite[DSS;][]{1995ASPC...77..179M}, and the galaxy evolution explorer \cite[GALEX;][]{2005ApJ...619L...1M}. Each region was examined at wide (3$^\circ$) and narrow (1\arcmin) fields of view to capture the surrounding environment and the immediate vicinity of the candidate.
These visual inspections were complemented by catalogue queries within a 1\arcmin\ radius around each candidate’s position. Sources from major astronomical databases (Vizier, SIMBAD, and NED) were retrieved to identify nearby objects and assess potential host associations.
 
Among the four candidates, only XT~111205 shows a clear association, being located within NGC~2264, a well-studied star-forming region in the Milky Way at a distance of 722~pc \citep{2023A&A...670A..37F}.
XT~111205 is also the only candidate projected onto the Galactic plane. At the distance of NGC~2264, its 0.5--7~keV luminosity is $L_X = 1.2^{+1.9}_{-0.7} \times 10^{29}$ erg s$^{-1}$, 
where we have used the flux from the blackbody model without Galactic absorption in \autoref{tab:t90z_fitpar}, considering the small inferred distance. 
For the remaining three transients, no convincing host associations were identified.

The luminosity and spectrum of XT~111205 are consistent with the properties of X-ray flares from dwarf stars, which typically have soft spectra and peak luminosities below $\sim 10^{30}\  \rm{erg\ s^{-1}}$, but reaching $\sim 10^{32}\  \rm{erg\ s^{-1}}$ in extreme cases \citep{Gudel2009,Robrade2010,Williams2014,Pye2015,DeLuca2020}. Such flares last $\sim 10^3 - 10^4$~s \citep{Pye2015}, longer than the 17~s duration observed for XT~111205, which may be attributed to the fact that the source is very faint and that we likely only detect the peak of the flare. The location and X-ray properties thus both favour the interpretation of XT~111205 as a stellar flare. Given the similar properties of the other three transients it is highly likely that they are also due to dwarf stars, although the lack of multi-wavelength counterparts and the locations away from the Galactic plane leave more ambiguity in the interpretation. 

If extragalactic in nature, these transients may alternatively be interpreted as SBOs. In particular, a SBO from a blue supergiant is expected to produce a $\sim 100$~s transient with a soft spectrum with $kT \sim 0.3$~keV, assuming a  non-relativistic spherically symmetric scenario \citep{Waxman2017}, which is broadly consistent with the properties of XT~040916 and XT~150429. Given the observed fluxes, a typical luminosity of a few times $10^{44}\ \rm{erg\ s^{-1}}$ for the SBOs would then imply a redshift $z \sim 3$ for these sources. Deeper follow-up observations are clearly needed in order to identify the stars and/or host galaxies associated with these transients. 

\subsection{Algorithm performance}
\label{sec:algoperformance}
Different approaches to transient searches with \textit{Chandra} data make different assumptions about source selection, exposure properties, and detection thresholds, which strongly affect completeness and contamination. Many previous studies rely on the CSC \citep{2010ApJS..189...37E,2024ApJS..274...22E} as their starting point \citep[e.g.][]{2022A&A...663A.168Q, Dillmann2025}, which ensures well-characterised sources but will miss faint transients and recent events (the latest catalogue is complete until the end of 2021, \citealt{2024ApJS..274...22E}).
Other works deliberately avoid catalogued sources, such as the single-frame search of \citet{2023MNRAS.523.2513Z}, which targets only events in regions without known point sources. The method we introduced is independent of pre-compiled source lists and applies only minimal pre-filtering (a source significance of 3$\sigma$). This makes it sensitive to catalogued and uncatalogued sources, across a wide range of sky regions, including the Galactic plane, and in fields that might otherwise be excluded.

Restrictions on exposure properties can also play an important role. For example, \citet{2019MNRAS.487.4721Y} limited their analysis to sources within 8\arcmin\ off-axis angle and divided exposures longer than 50~ks into shorter segments to mitigate background accumulation.
The method we introduced imposes no such cuts, allowing transients to be found in virtually all available observations. As a result, the algorithm is able to identify events on timescales ranging from single-frame spikes to longer-duration variability, although it is naturally restricted to sub-observation transients and primarily detects events lasting less than $\sim70\%$ of the exposure time (\autoref{fig:t90obsfrac}).

The detection thresholds for the sources and variability also vary widely between methods \citep{2019MNRAS.487.4721Y,Lin2022,2022A&A...663A.168Q,2023A&A...675A..44Q,2023MNRAS.523.2513Z, Dillmann2025}. Algorithms such as \texttt{glvary} in CIAO assign a variability index to each source based on deviations from a steady light curve \citep{Lin2022}, which is powerful for detecting persistent variability but less sensitive to isolated, short-duration events. Our method instead relies on \texttt{wavdetect} for initial source identification. 
Contamination in our search arises mainly from statistical fluctuations in faint sources, instrumental artefacts, or background variations; the conservative choices in \texttt{sigthresh} and the CuDiDi threshold are designed to minimise these false positives. 

Our subsequent cross-matching with external catalogues is also deliberately cautious, using a 3\arcsec\ radius, which further reduces contamination. 
We note, however, that this comes at the expense of a higher risks of excluding transients due to false associations (see Sect.~\ref{sec:selection}). This step could be improved in the future by using the positional uncertainties of all the individual sources instead of a fixed matching radius.  
Finally, while the algorithm can detect events as short as a single frame, the $T_{90}$ duration is derived from a sigmoid fit that is not always optimal for all light curves, especially for short transients embedded in very long exposures as discussed above.

\begin{figure}[htbp]
	 \resizebox{\hsize}{!}{\includegraphics[width=.9\textwidth]{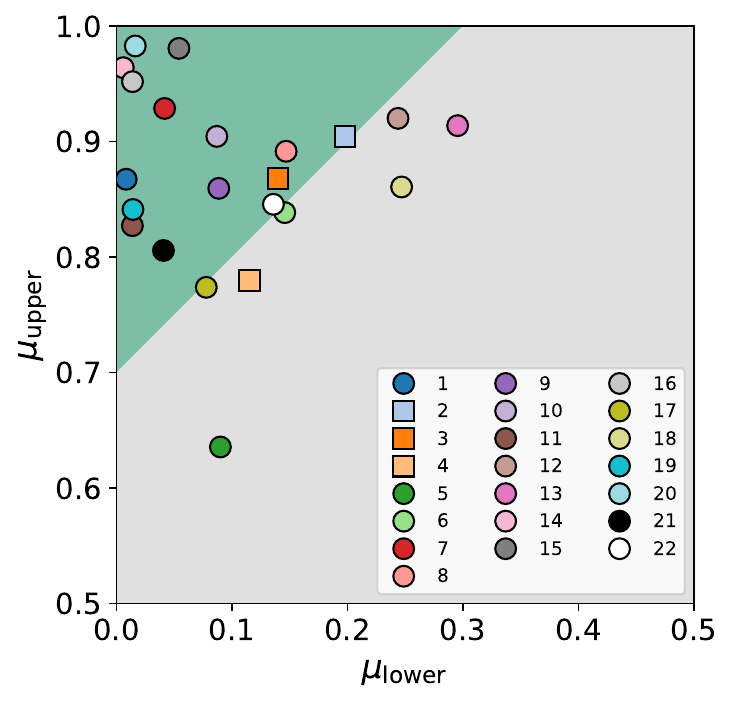}}
      \centering
	 \caption{CuDiDi diagram of the FXTs reported by \cite{2022A&A...663A.168Q} and \cite{2023A&A...675A..44Q}, labelled by numbers. The squares indicate the three FXTs that were excluded by our cross-matching criterion. Sources in the shaded green region are classified as transients according to the CuDiDi threshold. Further details about the FXTs are provided in \autoref{tab:QVcomparison}.
     }
     \label{fig:fxt_cudidi}
\end{figure}

The comparison with the transient samples of \citet{2022A&A...663A.168Q, 2023A&A...675A..44Q} demonstrates that our method is sensitive to previously reported transients, while applying stricter selection criteria. Out of the 22 published QV transients, our algorithm independently recovered 13, which were subsequently excluded from the catalogue (see \autoref{tab:QVcomparison} in \autoref{app:QVcomparison} for a detailed overview).
Out of the remaining nine sources, six were excluded by the CuDiDi threshold, while three were rejected due to associations with known objects in SIMBAD or NED (one of which was also below the CuDiDi threshold).

We plot all 22 FXTs on the CuDiDi diagram in \autoref{fig:fxt_cudidi}, which shows that the majority of sources that fall below the threshold do so by a very small margin,  illustrating that the CuDiDi cut is slightly more conservative than the method used in \citet{2022A&A...663A.168Q, 2023A&A...675A..44Q}.
 The source furthest away from the transient region in \autoref{fig:fxt_cudidi} is FXT~5, followed by FXT~13 and FXT~18. The light curve of FXT~5 (shown in \citealt[][Fig.~6]{2022A&A...663A.168Q}), exhibits two bursts of emission spanning $T_{90}=32.8~\mathrm{ks}$ of the total $51.7~\mathrm{ks}$ exposure, that is, $\sim63\%$ of the observation, which is at the upper end of our sample (\autoref{fig:t90obsfrac}). FXT~13 and 18 are both among the faintest of the FXT sample, with FXT~18 also having a $T_{90}=$ of 2.9~ks, which spans nearly half of the short $6.1~\mathrm{ks}$ exposure time.

The recovery of more than half of the QV transients provides an independent validation of our approach. 
The somewhat stricter criteria help guard against contamination from spurious detections, but can of course also miss real transients that fall just below the threshold. A future development of the CuDiDi classifier could therefore be to use a lower threshold, combined with additional tests of variability for sources that fall just above the boundary.
Further analysis of the golden sample, including host galaxy identifications, will be necessary to determine whether the transients uncovered in this search differ from those reported in previous studies.

\section{Summary and conclusions}\label{sec:summary}
We have presented a search for sub-observation transients in the full public \textit{Chandra} ACIS archive, covering 20,212 observations obtained before 2024-12-31. Our transient detection pipeline integrates automated source detection, light-curve extraction, diagnostic flagging, and the novel CuDiDi classifier, followed by visual screening to minimise contamination. Previously known sources were removed by cross-matching with other catalogues, ensuring that the resulting sample comprises single-occurrence transients, such as cataclysmic events, and new detections of flaring objects. The main outcomes of this work are summarised below.

\begin{enumerate}
    \item From 1420 initial detections, we constructed a golden sample of 765 high-confidence transients, suitable for follow-up studies.
 
    \item The sample spans a wide range of temporal properties, with $T_{90}$ values covering five orders of magnitude. While most events lasted thousands of seconds, we also identified a tail of very fast events, with durations $\lesssim 30$~s. The transients are also spectrally diverse, with a wide range of hardness ratios. 
  
    \item  A more detailed investigation of four fast candidates with durations $\lesssim 30$~s showed that their spectra are very soft. The candidates can be characterised as blackbodies with $kT$ in the range $\sim 0.06-0.4$~keV. One of them (XT~111205) is associated with a star-forming region in the Galactic plane, implying a low luminosity of $ \sim 10^{29}\ \rm{erg\ s^{-1}\ cm^{-2}}$ and making it consistent with a flare from a dwarf star. The other three candidates lack multi-wavelength/host galaxy associations, but their properties are similar to those of XT~111205, which makes it likely that they are also flaring stars. 

    \item A comparison with the transients reported in \citet{2022A&A...663A.168Q,2023A&A...675A..44Q} showed that CuDiDi performed well in terms of identifying transients, while the threshold we used here imposed a somewhat stricter variability criterion.
 \end{enumerate}

Future work will include further analysis  of the transient catalogue, including host galaxy identification and more detailed studies of the X-ray properties. The CuDiDi classifier will also be explored further, including its potential use as a tool for identifying repeating sources. 

\section*{Data availability}
The following data products are available on Zenodo: 
the final candidate catalogue: \url{https://doi.org/10.5281/zenodo.16356970},
the complete list of observation IDs analysed: \url{https://doi.org/10.5281/zenodo.17099991},
the types of objects excluded in the SIMBAD and NED searches: \url{https://doi.org/10.5281/zenodo.16759113}.

\begin{acknowledgements}
  The authors thank Dennis Alp for his consultations and fruitful discussions. This research was supported by the Knut \& Alice Wallenberg foundation. 

  This research has used the SIMBAD database, operated at CDS, Strasbourg, France.
  This research has used the NASA/IPAC Extragalactic Database (NED), which is funded by the National Aeronautics and Space Administration and operated by the California Institute of Technology.
  This research has made use of Aladin sky atlas developed at CDS, Strasbourg Observatory, France.
 This work has used the following software: 
Astropy 5.3.4 \citep{astropy:2013, astropy:2018, astropy:2022},
ds9 8.4.1, CIAO 4.15.2 with CALDB 4.10.7 \citep{2006SPIE.6270E..1VF},
Heasoft ftools 22Aug2023$\_$V6.32.1 \citep{1999ascl.soft12002B}
Matplotlib 3.8.1 \citep{Hunter:2007}, 
Numpy 1.26.0 \citep{harris2020array}, 
Pandas 2.2.0 \citep{mckinney-proc-scipy-2010}
.
\end{acknowledgements}


\bibliographystyle{aa}
\bibliography{bibliography}


\appendix
\twocolumn
\section{Comparison with Quirola-Vásquez et al. (2022, 2023)}
\label{app:QVcomparison}

Table~\ref{tab:QVcomparison} lists the transient candidates reported by \citet{2022A&A...663A.168Q, 2023A&A...675A..44Q}, their \textit{Chandra} observation IDs, positions, and whether the sources were recovered in our analysis. For non-recovered sources, the reason for exclusion is indicated.

\begin{table*}[!htbp]
\caption{Comparison with transient candidates reported by \citet{2022A&A...663A.168Q, 2023A&A...675A..44Q}.}
\label{tab:QVcomparison}
\centering
\begin{tabular}{lcrrl}
	\hline\hline
	FXT & Obs. ID & RA (deg) & Dec (deg) & Status / Reason (this work) \\ 
	\hline
        \multicolumn{5}{c}{Quirola-Vásquez et al. (2022)} \\
        \hline
	1 & 803  & 186.38125 & 13.06607 & Recovered (removed from catalogue) \\
	2 & 2025  & 167.86792 & 55.67253 & Known HMXB (SIMBAD), (above CuDiDi threshold) \\
	3 & 8490  & 201.24329 & $-$43.04060 & Planetary Nebula (SIMBAD/NED), (above CuDiDi threshold) \\
	4 & 9546  & 211.25113 & 53.65706 & Known HMXB (SIMBAD), below CuDiDi threshold \\
	5 & 9548 & 170.07296 & 12.97189 & Below CuDiDi threshold \\
	6 & 14904 & 345.49250 & 15.94871 & Below CuDiDi threshold \\
	7 & 4062 & 76.77817 & $-$31.86980 & Recovered (removed from catalogue) \\
	8 & 5885 & 318.12646 & $-$63.49914 & Recovered (removed from catalogue) \\
	9 & 9841 & 175.00504 & $-$31.91743 & Recovered (removed from catalogue) \\
	10 & 12264 & 90.00450 & $-$52.71501 & Recovered (removed from catalogue) \\
        11 & 12884 & 212.12063 & $-$27.05784 & Recovered (removed from catalogue) \\
        12 & 13454 & 15.93558 & $-$21.81272  & Below CuDiDi threshold  \\
        13 & 15113 & 45.26725 & $-$77.88095  & Below CuDiDi threshold \\
        14 & 16454 & 53.16158 & $-$27.85940  & Recovered (removed from catalogue) \\
        \hline
        \multicolumn{5}{c}{Quirola-Vásquez et al. (2023)} \\
        \hline
        15 & 16093 & 233.73496 & 23.46849  & Recovered (removed from catalogue) \\
        16 & 16453 & 53.07672 & $-$27.87345  & Recovered (removed from catalogue) \\
        17 & 18715 & 40.82972 & 32.32390 & Below CuDiDi threshold \\
        18 & 19310 & 36.71489 & $-$1.08317 & Below CuDiDi threshold \\
        19 & 20635 & 356.26437 & $-$42.64494  & Recovered (removed from catalogue) \\
        20 & 21831 & 207.34711 & 26.58421  & Recovered (removed from catalogue) \\
        21 & 23103 & 50.47516 & 41.24704 & Recovered (removed from catalogue) \\
        22 & 24604 & 207.23523 & 26.66230 & Recovered (removed from catalogue) \\
	\hline
\end{tabular}
\tablefoot{Listed are the QV designation (FXT, followed by a number), \textit{Chandra} observation ID, position, and whether the source was recovered in our analysis. For non-recovered sources, the reason is indicated.}
\end{table*}

\section{Impact of background}
\label{app:bkg}

The CuDiDi method used in this work is based on total cumulative counts without background subtraction. 
This is motivated by the generally very low background in \textit{Chandra}, combined with the use of short 1~s bins to improve sensitivity to short timescale variability. Background subtraction in finely time resolved bins leads to bins with negative counts, which in turn results in a cumulative distribution that does not increase monotonously, for which the  CuDiDi discriminator was not designed. 

To explore the impact of background we take 10 transient candidates that were identified in observations with exposure times >100~ks, considering that the background is expected to be more important in long observations. We bin the background-subtracted light curves to 5~ks and set the count rate in any intervals that have negative counts to 0 when constructing the cumulative distributions. The results are shown in \autoref{fig:bkg_cudidi}, where we also compare with the results of using 5~ks bins without background subtraction, and the original 1~s bins. As can be seen, the background subtraction has a small effect on the results and do not shift the points in any systematic way. In fact, \autoref{fig:bkg_cudidi} shows that changing the bins form 1~s to 5~ks generally has a larger impact on the positions in the CuDiDi diagram than the background subtraction.

These results indicate that the transient selection based on total cumulative counts is not strongly impacted by background. However, extending this method to other telescopes with higher background would most likely require some adjustments of the method. This could involve using wider background-subtracted bins as shown here, or simply adjusting the selection criteria in the CuDiDi plane to account for different background levels.  

\begin{figure}[htbp]
	 \resizebox{\hsize}{!}{\includegraphics[width=\textwidth]{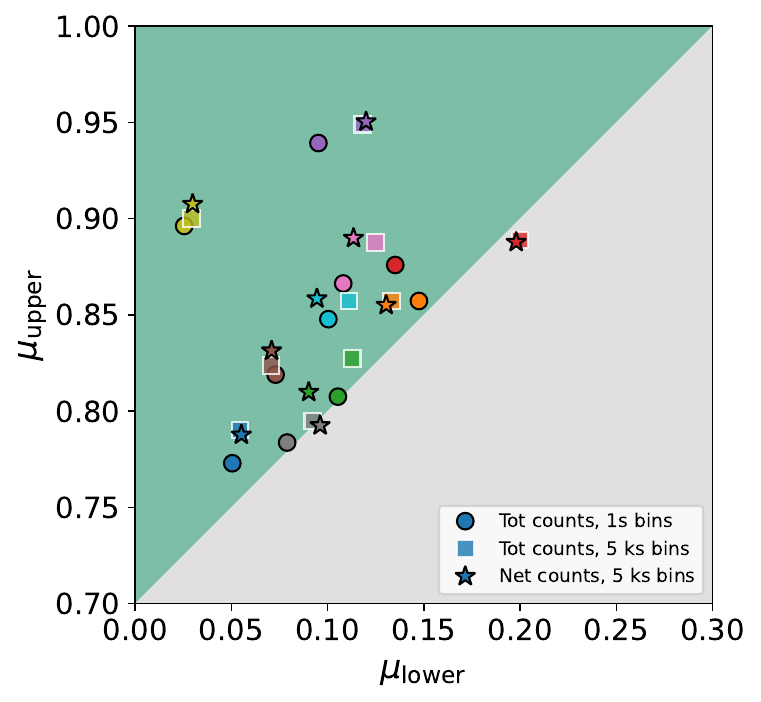}}
      \centering
	 \caption{CuDiDi diagram of 10 transit candidates that were identified in observations with exposure times >100~ks. Each transient is shown with a different colour, where the different symbols correspond to different choices of binning and background subtraction as indicated in the legend. Only the upper left corner of the CuDiDi diagram is shown in order to highlight the differences between the different points.
     }
     \label{fig:bkg_cudidi}
\end{figure}

\twocolumn

\end{document}